\documentclass[aps, prl, twocolumn, superscriptaddress, amsmath, showpacs, tightenlines, footinbib, longbibliography]{revtex4-1}
\usepackage[colorlinks, breaklinks, citecolor=blue, linkcolor=blue,  urlcolor={blue}]{hyperref}
\usepackage{breakurl}
\usepackage{amsfonts}
\usepackage{amsmath}
\usepackage{amssymb}
\usepackage{mathrsfs}
\usepackage{epsfig}
\usepackage{graphicx}
\usepackage{mathtools}
\usepackage{bm}
\usepackage{natbib}

\usepackage{colortbl}
\definecolor{mygray}{gray}{.9}
\definecolor{darkblue}{rgb}{1,1,.70}
\definecolor{lightblue}{rgb}{1,1,.90}
\newcommand{\Tin}[1]{{\color{black} #1}}

\begin{document}

\title{Quantum Magnetometer with Dual-Coupling Optomechanics}

\author{Gui-Lei Zhu}
\affiliation{School of Physics, Huazhong University of Science and Technology, Wuhan 430074, China}
\author{Jing Liu}
\affiliation{School of Physics, Huazhong University of Science and Technology, Wuhan 430074, China}
\affiliation{MOE Key Laboratory of Fundamental Physical Quantities Measurement and PGMF, Huazhong University of Science and Technology, Wuhan 430074, China}
\author{Ying Wu}
\affiliation{School of Physics, Huazhong University of Science and Technology, Wuhan 430074, China}

\author{Xin-You L\"{u}}
\email{xinyoulu@hust.edu.cn}
\affiliation{School of Physics, Huazhong University of Science and Technology, Wuhan 430074, China}
\date{\today}

\begin{abstract}
An experimentally feasible magnetometer based on a dual-coupling optomechanical system is proposed, where the radiation-pressure coupling transduces the magnetic signal to the optical phase, and the quadratic optomechanical interaction induces a periodic squeezing effect. The latter not only amplifies the signal to be measured, but also accelerates the signal transducing rate characterized by an experimentally observable phase accumulation efficiency. In the vicinity of opto-mechanical decoupled time, the ultimate bound to the estimability of magnetic signal is proportional to $\exp(-6r)$, and then the optimized accuracy of estimation can be enhanced nearly 3 orders with a controllable squeezing parameter $r<1$. Moreover, our proposal is robust against the mechanical thermal noise, and the sensitivity of a specific measurement can reach to the order of $10^{-17}{\rm T/\sqrt{Hz}}$ in the presence of dissipations and without ground state cooling of mechanical oscillator. Our proposal fundamentally broadens the fields of quantum metrology and cavity optomechanics, with potential application for on-chip magnetic detection with high precision.

\end{abstract}

\maketitle
Ultrasensitive magnetic detection has contributed immensely to a wide range of scientific areas from fundamental physics to advanced technologies, such as geological exploration, aerospace\,\cite{Bennett2021}, biomedical imaging and diagnostics\,\cite{Hamalainen1993,Lee2015,Murzin2020}. Over the past few decades, various magnetometers have been developed, including the superconducting quantum interference devices (SQUID)  based on superconducting effects\,\cite{Jaklevic1964,Erne1976,Kleiner2004}, spin-exchange relaxation-free atomic magnetometers\,\cite{Allred2002,Kominis2003,Savukov2005,Xia2006}, NV center magnetometers\,\cite{Taylor2008,Maze2008} and Hall-effect sensors\,\cite{BENDING1999}. Normally, they require the elaborated operating conditions, such as the associated denoising technology and/or the complex signal read-out schemes\,\cite{robbes2006}, which reduces their capability of on-chip integration.

Cavity optomechanical system (OMS)\,\cite{Aspelmeyer2014Rev,Kippenberg2008,Aspelmeyer2012,Meystre2013} offers an alternative platform for the precision measurements of mass\,\cite{zhukadi2012,Lin2017,Shang-Wu2019}, weak forces\,\cite{Clerk2010,Tsang2010,Pontin2014,Armata2017,Qvarfort2018}, and magnetic fields\,\cite{Forstner201201,ForstnerAdvanced2014,Yu2016,Wu2017,Zhu2017,Li2018}. Particularly, optomechanical magnetometer with high sensitivity has excellent quality of on-chip integration~\cite{Bei-Bei2021}. Recently, the YIG sphere-based optomechanical magnetometer has been experimentally demonstrated in Ref.\,\cite{colombano2020}, which have attained extremely low sensitivity values. In such magnetometers, the existence of Joule and Villari effects of magnetostrictive transducer\,\cite{olabi2008}, allows one to directly extract magnetic information by reading out the optical frequency shift (or transmission spectrum). Moreover, quantum metrology\,\cite{Giovannetti2004,Giovannetti2006,Giovannetti2011,Dowling2015} points out that the quantum squeezing or entanglement\,\cite{Caves1981,Ma2011,Baumgratz2016,Degen2017,Engelsen2017,Nagata2007,Israel2014,Luo2017} could improve the accuracy of parameter estimation in physical systems from the shot-noise limit to the Heisenberg limit, i.e., the optimal precision scales from $1/\sqrt{N}$ to $1/N$ with $N$ being the number of resources employed in the measurements. This has stimulated enormous interests in exploiting quantum resources in the atoms (or spins)\,\cite{Jones2009,Tanaka2015,Komar2014,Hou2020} and optical systems\,\cite{Holland1993,Boto2000,Anisimov2010,Joo2011}  for high-precision physical quantity measurements.

By applying quantum metrology to the detection of a static magnetic field, here we propose a quantum magnetometer based on a dual-coupling OMS, which has a periodic decoupling behavior between the optical and mechanical modes. The OMS supports two optical modes coupled simultaneously to the same mechanical mode, with radiation-pressure and quadratic optomechanical interactions, respectively\,\cite{thompson2008,Sankey2010,Bhattacharya2008,Zhu2018}. The radiation-pressure coupling acts as a signal transducer, encoding the magnetic signal received by the mechanical oscillator into the optical phase. The quadratic optomechanical interaction amplifies both the signal to be measured and the signal transducing rate via inducing a periodic squeezing effect on the mechanical oscillator, whose maximum squeezing strength is determined by a controllable squeezing parameter $r$. By performing a homodyne detection on the optical phase within a wide time window around the first decoupled time $\tau_1$, the magnetic signal could be estimated with high precision.

To qualitatively characterize the precision of magnetometer, we define a displaced phase accumulation efficiency (PAE) that is  experimentally observable via state tomography technique. By presenting the {\it exponentially} increased quantum Fisher information (QFI), i.e., $\mathcal{F}_q(\tau_1)\propto e^{12 r}$, we quantitatively demonstrate that the fundamental bound of measurement precision can be dramatically reduced even with a small squeezing parameter $r$. The periodic opto-mechanical decoupling makes the classical Fisher information (CFI) robust against the mechanical environment at the decoupled time, which in turn allows the sensitivity of a specific measurement to saturate the fundamental bound and reach to the order of $10^{-17}{\rm T/\sqrt{Hz}}$ in the presence of system dissipations. Moreover, the sensitivity to the order of $\sim10^{-15}{\rm T/\sqrt{Hz}}$ is predicted even in the case of the thermal phonon number $\bar{n}_{\rm th}\sim10^3$. Our work establishes a connection between quantum metrology and dual-coupling optomechanics, which is suitable for detecting various fields that linearly interact with the mechanical oscillator.
%================================================
\begin{figure}
\includegraphics[width=7.2cm]{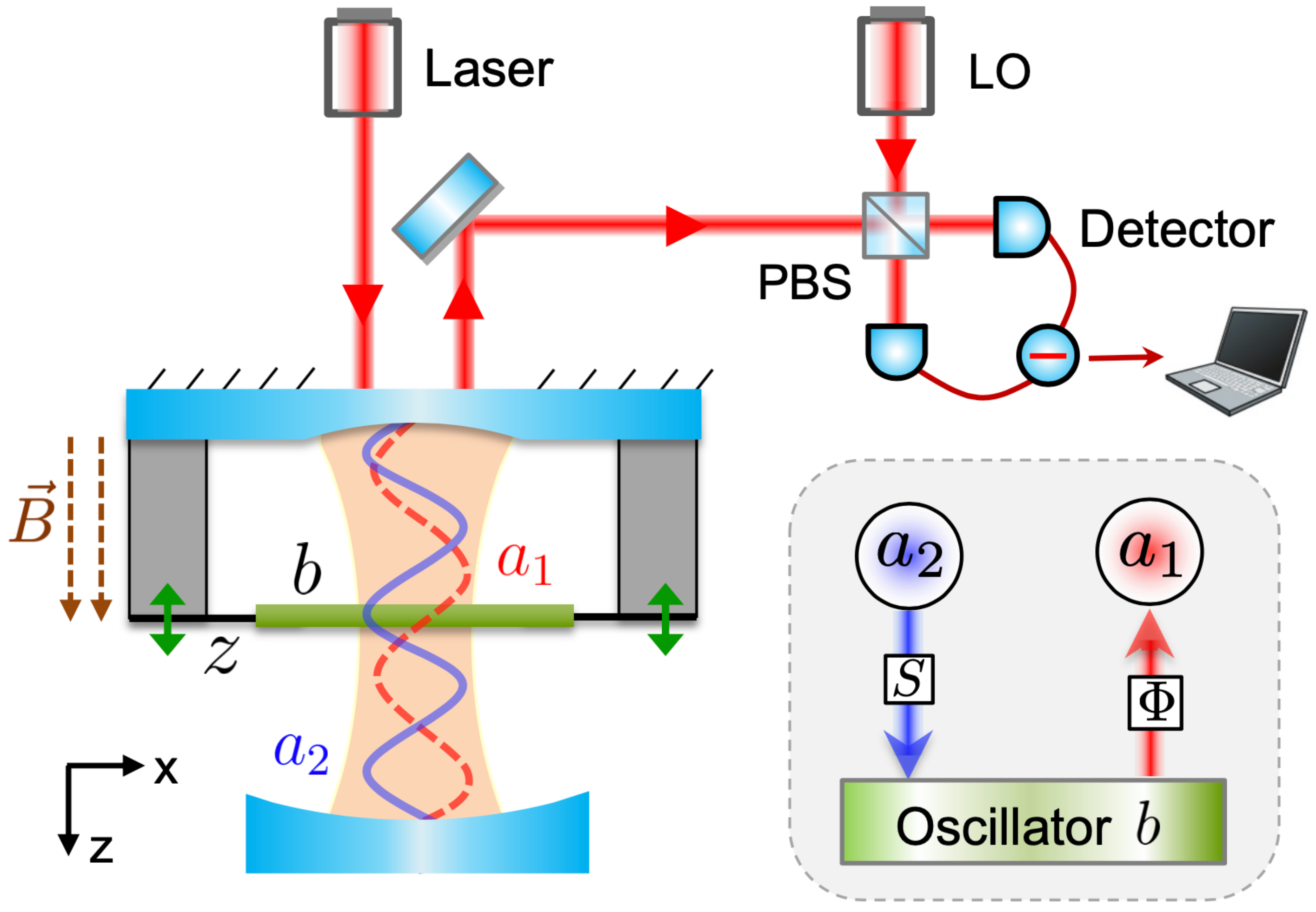}
\caption{Schematic illustration of quantum magnetometer based on a dual-coupling OMS with ``membrane-in-the-middle'' configuration. The middle SiN membrane (green), acting as the mechanical mode $b$, is welded to the ends of magnetostrictive Terfenol-D rods (gray)\,\cite{Hong2013}. On application of a static magnetic field, the Terfenol-D rods expands, which displaces the equilibrium position of the mechanical mode, and then the magnetic signal is encoded into the phase $\Phi$ of optical mode $a_1$ via radiation-pressure coupling. Besides, cavity mode $a_2$ offers the periodic squeezing for mode $b$ via the quadratic optomechanical interaction [see the shaded area]. The optical phase is detected by the balanced homodyne detection scheme with a local oscillator (LO) pulse.}
\label{fig1}
\end{figure}
%================================================

\emph{System and periodic mechanical squeezing.---} We consider a dual-coupling optomechanical system depicted in Fig.\,\ref{fig1} with Hamiltonian
\begin{align}
H&=\hbar\omega_1a^{\dagger}_1a_1+\hbar\omega_2a^{\dagger}_2a_2+\hbar\omega_m b^{\dagger}b-\hbar\lambda_1a^{\dagger}_1a_1(b^{\dagger}+b)\nonumber
\\
&\,\,\,\,\,\,\,-\hbar\lambda_2a^{\dagger}_2a_2(b^{\dagger}+b)^2+B_zc_{\rm act}z,\label{H_original}
\end{align}
where $a_j$ $(j=1,2)$ and $b$ are the annihilation operators of the cavity mode with frequency $\omega_j$ and the mechanical mode with frequency $\omega_m$, respectively. The mechanical membrane is placed at a node (antinode) of cavity mode $a_1$ ($a_2$), and then the fourth (fifth) term in Eq.\,(\ref{H_original}) describes the radiation-pressure (quadratic optomechanical) interaction between modes $a_1$ ($a_2$) and $b$ with strength $\lambda_1$ ($\lambda_2$). In the presence of a static magnetic field ${B_z}$ (along $z$ direction), the field-sensitive Terfenol-D expands, which leads to the change of the equilibrium position of the mechanical oscillator, thus generating an effective magnetic potential on the Hamiltonian, i.e., the last term of Equation\,(\ref{H_original})\,\cite{SM}. The mechanical motion modulates the optical cavity field via the nonlinear radiation-pressure coupling. Meanwhile, the phase shift of the mechanical motions encoded with magnetic signal is transferred to the optical field. By reading out the phase shift of optical field via homodyne detection, we can extract
the original magnetic information. Here $z=\sqrt{\hbar/(2m\omega_m)}(b^{\dag}+b)$ is the position operator of the mechanical oscillator with mass $m$, and $c_{\rm act}=m\omega_m^2L{\alpha_{\rm mag}}/{E}$ is the magnetic actuation constant charactering how well the magnetic field is converted into a force applied on the oscillator. The symbol $L$ denotes the length of Terfenol-D rods, $\alpha_{\rm mag}$ is magnetostrictive coefficient and $E$ is the Young's modulus\,\cite{SM}.
%================================================
\begin{figure}
	\includegraphics[width=8.5cm]{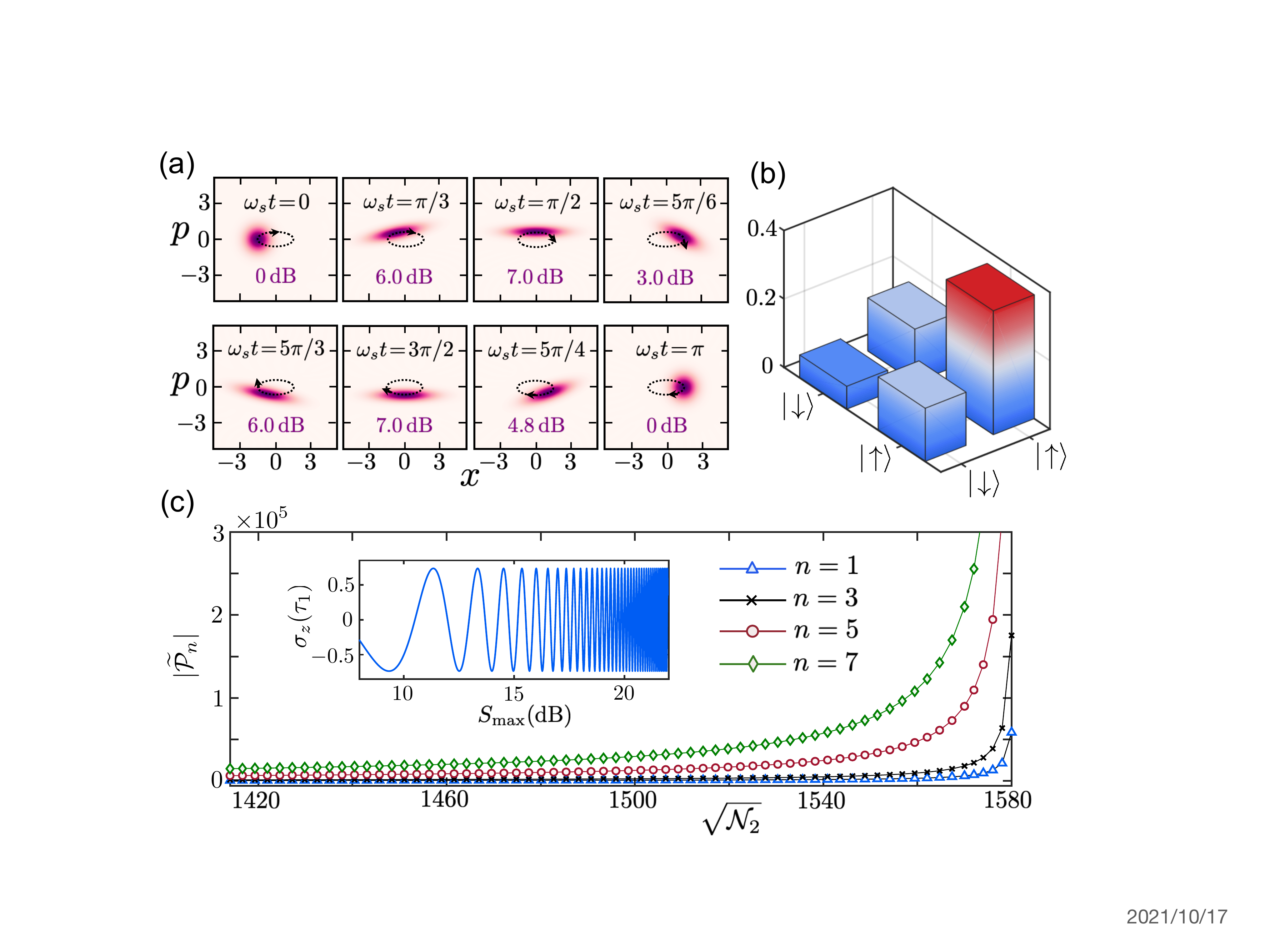}
	\caption{(a) Wigner functions of the mechanical mode within 2$T$ given by $a_1$ in vaccum state $|0\rangle$,  $\omega_m=1$ and {$\sqrt{\mathcal{N}_2}=1414$}. The instantaneous squeezing degrees are marked. (b) Tomography of the state $|\Psi(\tau_1)\rangle$ projected on the subsapce $\{|\!\!\downarrow\rangle,|\!\!\uparrow\rangle\}$ with $l=3$ and $\sqrt{\mathcal{N}_2}\!=\!1500$. The dependence of $\sigma_z(\tau_1)$ on the maximum squeezing degree $S_{\rm max}$ for $l=1$ and $f=\omega_m$ is shown in the insert of (c). The main panel of (c) shows the absolute value of displaced PAE $|\widetilde{\mathcal{P}}_n|$ as a function of $\sqrt{\mathcal{N}_2}$ for several values of $n$. The parameters are $ \omega_m=2\pi\times134 \,{\rm kHz}, f=0.01\omega_m$,  $\alpha=1$, $\lambda_1=0.01\omega_m$, and $\lambda_2=10^{-7}\omega_m$.}
	\label{fig2}
\end{figure}
%================================================

By considering the ancillary mode $a_2$ in the coherent state $|\xi\rangle$, the number operator $a_2^{\dagger}a_2$ can be approximately replaced by an algebraic number $\mathcal{N}_2=|\xi|^2$ in the case of $\mathcal{N}_2\gg1$\,\cite{SM}. Assuming the modes $a_1$ and $b$ are initially in the coherent state $|\Psi(0)\rangle=|\alpha\rangle\otimes|\beta\rangle$ with $\beta=\beta_{\rm Re}+i\beta_{\rm Im}$, the instantaneous state of system is given by\,\cite{SM}
\begin{align}\label{Psi_t}
&\!\!\!|\Psi(t)\rangle\!\!=\!e^{-{|\alpha|^2}/{2}}\sum_{n=0}^{\infty}\frac{\alpha^n}{\sqrt{n!}}{\rm exp}\left[i\eta^2(\omega_st-\sin\omega_st)\right]\nonumber
\\
&\!\!\!\times\!{\rm exp}\{{i\eta[\beta_{\rm Re}\sin\omega_st\, e^{-r}\!\!-\!\!\beta_{\rm Im}(\cos\omega_st\!-\!\!1)e^{r}]}\}|n\rangle|\varphi_s(t)\rangle,\!
\end{align}
where $\eta=\lambda_s n/\omega_s-f_s/\omega_s$ with $\lambda_{s}=\lambda_1 e^{r}$, $\omega_s=\omega_m e^{-2r}$,  $f_s=fe^r=B_zc_{\rm act}\sqrt{1/(2m\hbar \omega_m)}e^r$, $r=-(1/4)\ln(1-4\lambda_2\mathcal{N}_2/\omega_m)$, and a rotating frame with ${\rm exp}(-i\omega_1a_1^{\dag}a_1/\omega_s)$ was adopted. The mechanical state reads $|\varphi_s(t)\rangle=S^{\dagger}(r)S(r')|\varphi_n(t)\rangle$ with the defined squeezing operator $S(r)=\exp[r(b^2-b^{\dagger2})/2]$ and the squeezing parameter $r'=re^{-2i\omega_st}$. Here $|\varphi_n(t)\rangle=|e^{-i\omega_st}\beta+\eta\bar{\mu}\rangle$ is a displaced coherent state with $\bar{\mu}=(1-e^{-i\omega_st})(\cosh r-e^{-i\omega_st}\sinh r)$. The expression of $|\varphi_s(t)\rangle$ clearly shows that the mechanical mode is dynamically squeezed with the period $T=\pi/\omega_s$, whose squeezing degree of quadrature $X=1/\sqrt{2}(b+b^{\dagger})$ is defined by $S(t)=10\log_{10}(\delta X^2(t)/\delta X_{\rm min}^2){\rm dB}$. The maximum squeezing degree $S_{\rm max}=10\log_{10}(e^{4r}){\rm dB}$ occurs at $T/2$ with the state $S^{\dagger}(2r)|\beta\rangle$\,\cite{SM}. This periodic squeezing effect on the mechanical oscillator is induced by the quadratic optomechanical coupling, and can be qualitatively presented in the case of $\alpha=\eta=0$ [see Fig.\,\ref{fig2}(a)].

\emph{Periodic-squeezing-enhanced phase accumulation efficiency.---}
As shown in Eq.\,(\ref{Psi_t}), the magnetic signal is transduced into the optical phase during the evolution of system via the radiation-pressure interaction. Interestingly, at time $\tau_m\!=\!2m\pi/\omega_s$ ($m=1,2,....$),
Eq.\,({\ref{Psi_t}}) can be reduced to\,\cite{SM}
\begin{align}
|\Psi(\tau_m)\rangle=e^{-{|\alpha|^2}/{2}}\sum_{n=0}^{\infty}\frac{\alpha^n}{\sqrt{n!}}e^{i\Phi_n(\tau_m)}|n\rangle|\beta\rangle,
\end{align}
which demonstrates that the optical and mechanical modes are periodically decoupled, meanwhile the magnetic signal to be measured is periodically encoded into the accumulated optical phase $\Phi_n(\tau_m)=2m\pi (\lambda_s n/\omega_s-f_s/\omega_s)^2$ for a Fock state $|n\rangle$.  Here the opto-mechanical decoupled period is double of the period of dynamical squeezing, i.e., $\tau_1=2\pi/\omega_s=2T$, and hence the mechanical squeezing also disappears at the decoupled time $\tau_m$.

The above unique property allows us to estimate the magnetic field $B_z$ by performing a homodyne detection [see Fig.\,\ref{fig1}] on the optical mode $a_1$ at the first decoupled time $\tau_1$. To qualitatively describe the detection precision, here we define a displaced PAE
\begin{align} \widetilde{\mathcal{P}}_n=\frac{\Phi_n(\tau_1)-\Phi_0(\tau_1)}{\tau_1}=\frac{{\lambda_1n}}{\omega_m}{{(\lambda}_1n-2{f})}{e^{4r}}.
\end{align}
Obviously, the larger $\widetilde{\mathcal{P}}_n$, i.e., the faster phase accumulation rate on the optical Fock state $|n\rangle$, the higher measurement accuracy of magnetometer should be obtained. As shown in Fig.\,\ref{fig2}(c), the $\widetilde{\mathcal{P}}_n$ is exponentially enhanced by increasing the coherent amplitude $\sqrt{\mathcal{N}_2}$ of the ancillary mode $a_2$. This enhancement originally comes from the periodic squeezing of mechanical mode during one opto-mechanical decoupled period [see Fig.\,\ref{fig2}(a)]. The system Hamiltonian in the squeezed frame shown in Eq.\,(S17) of supplementary material~\cite{SM} clearly demonstrates that the mechanical squeezing effect not only accelerates the signal tranducing rate from phonon to photon (i.e., the radiation-pressure interaction $\lambda_sa^{\dagger}_1a_1(b+b^{\dagger})$)~\cite{Xin-You2015,Lemonde2016}, but also amplifies the magnetic signal to be measured (i.e., the magnetic potential $f_s(b^{\dag}+b)$).
%================================================
\begin{figure}
	\includegraphics[width=8cm]{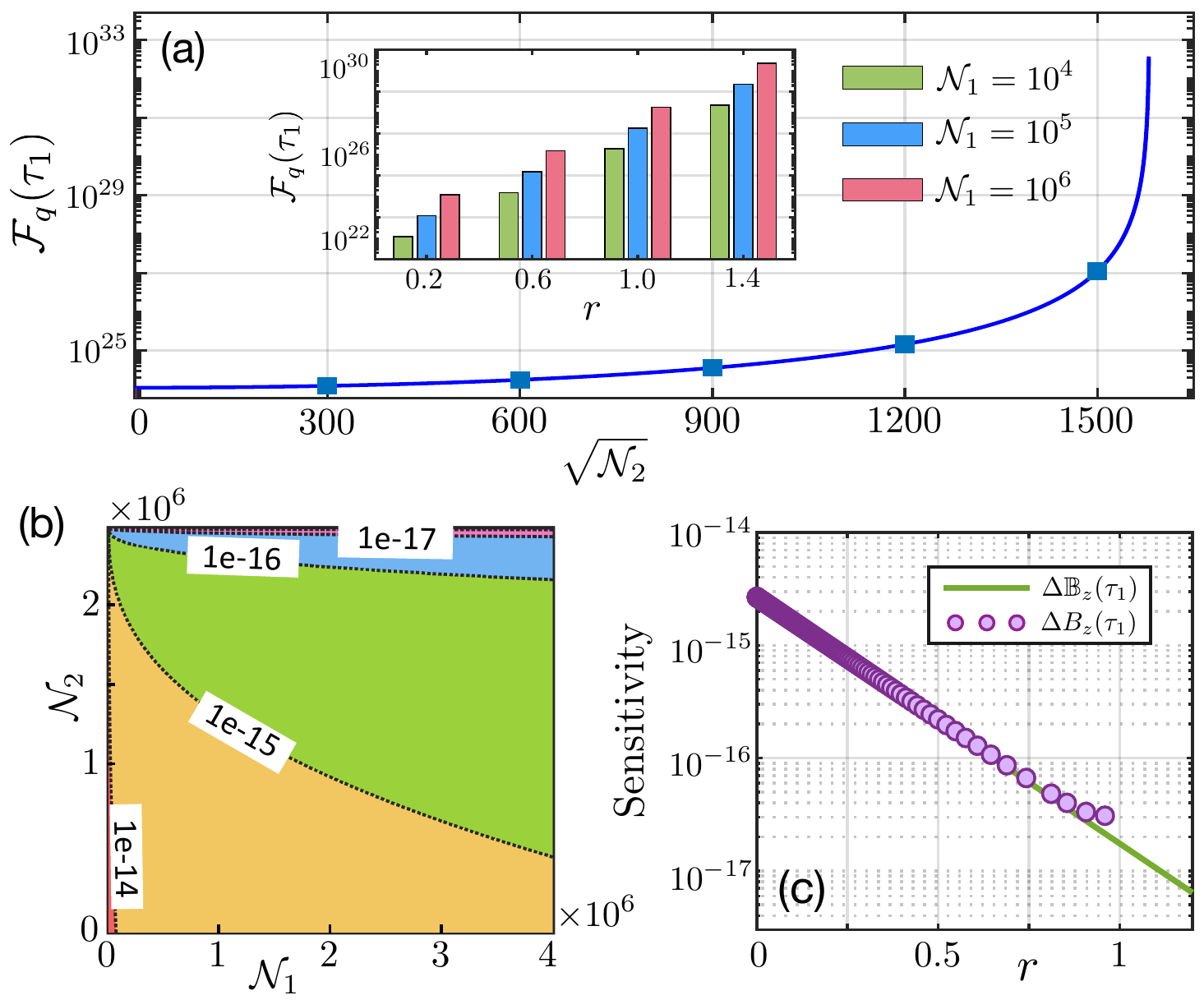}
	\caption{(a) The QFI $\mathcal{F}_{q}(\tau_1)$ versus $\sqrt{\mathcal{N}_2}$ (main) and $r$ (inset). (b) Contourf plot of the optimal sensitivity $\Delta \mathbb{B}_z (\tau_1)$ in units of ${\rm T/\sqrt{Hz}}$ as functions of $\mathcal{N}_1$ and $\mathcal{N}_2$. (c) $\Delta \mathbb{B}_z (\tau_1)$  (solid curve) and the specific sensitivity $\Delta B_z (\tau_1)$ in the presence of dissipations (circles) versus $r$ for $\mathcal{N}_1=10^6$. Here we have chosen $m=4\times 10^{-8}{\rm g},$ $L=630\,\mu$m, $\alpha_{\rm mag}=5\times 10^8{\rm NT^{-1}m^{-1}}$, $E=30{\rm \,GPa}$,  $\kappa/\omega_m=0.01, \gamma/\omega_m=0.001$, $\bar{n}_{\rm th}=10$ and other parameters are same as Fig.\,2.}
	\label{fig3}
\end{figure}
%================================================

More importantly, this well-defined PAE is experimentally observable via state tomography in the proper basis vectors. Specifically, expanding the system state $|\Psi(\tau_1)\rangle$ in the subspace $\{|\!\!\downarrow\rangle,|\!\!\uparrow\rangle\}$ with basis vectors $|\!\downarrow\rangle:=(1/\sqrt{2})(|0\rangle-|l\rangle)$ and $|\!\uparrow\rangle:=(1/\sqrt{2})(|0\rangle+|l\rangle)$ ($l=1,2,....$), we obtain four elements of density matrix\,\cite{SM}, as shown in Fig.\,\ref{fig2}(b). The difference between two diagonal elements is denoted by $\sigma_z(\tau_1)=\rho_{\uparrow\uparrow}(\tau_1)-\rho_{\downarrow\downarrow}(\tau_1)=2e^{-|\alpha|^2}(\alpha^l/\sqrt{l!})\cos(\Delta\Phi_l(\tau_1))$, where $\Delta\Phi_l(\tau_1)=\Phi_l(\tau_1)-\Phi_0(\tau_1)$. Then one can easily obtain the values of phase difference $\Delta \Phi_l(\tau_1)$ by directly measuring $\rho_{\downarrow\downarrow}(\tau_1)$ and $\rho_{\uparrow\uparrow}(\tau_1)$, which ultimately leads to $\widetilde{\mathcal{P}}_l=\Delta \Phi_l(\tau_1)/\tau_1$ being experimentally observable. Moreover, the oscillation with increasing frequency of $\sigma_z(\tau_1)$, shown in the insert of Fig.\,\ref{fig2}(c), is another evidence for the enhanced $\widetilde{\mathcal{P}}_n$ along with increasing the mechanical squeezing.

\emph{Quantum and classical Fisher information.---}
From a quantitative point of view, the fundamental bound to the sensitivity and the measurement-specific sensitivity of the proposed magnetometer are respectively decided by the QFI $\mathcal{F}_q$ and CFI $\mathcal{F}_c$ based on the Cram\'{e}r-Rao inequality ${\Delta}B_z\ge 1/\sqrt{M\mathcal{F}_j}$ $(j=q, c)$, where $\Delta B_z$ is the standard deviation with respect to an unbiased estimator $(B_z)_{\rm est}$, and $M$ is the number of repetition of the experiments\,\cite{Braunstein1994}.

Considering system state $|\Psi(t)\rangle$, the QFI with respect to the parameter $B_z$ reads $\mathcal{F}_{q}(t)\!=\!4(\langle \partial_{B_z} \Psi(t)|\partial_{B_z}\Psi(t)\rangle\!-\!\left|\langle \Psi(t)|\partial_{B_z}\Psi(t)\rangle\right|^2)$, where $\partial_{B_z}=\partial/\partial_{B_z}$\,\cite{paris2009,toth2014,liu201901,Lu2021,liu2022optimal}.
At the first decoupled time $\tau_1$, the QFI reduces to\,\cite{SM}
\begin{align}\label{QFI}
\mathcal{F}_{q}(\tau_1)&=\frac{32\pi^2m\lambda_1^2L^2\alpha^2_{\rm mag}}{\hbar \omega_mE^2}\mathcal{N}_1e^{12r},\nonumber
\\
&=\frac{32\pi^2m\lambda_1^2L^2\alpha^2_{\rm mag}}{\hbar \omega_mE^2}\frac{\mathcal{N}_1}{[1-(4\lambda_2/\omega_m)\mathcal{N}_2]^3}.
\end{align}
where $\mathcal{N}_1=|\alpha|^2$ is the mean photon number of cavity mode $a_1$. It can be seen that the QFI is exponentially enhanced with power of $12r$, and hence increasing a small value of $r$ by changing $\mathcal{N}_2$ can give rise to a large enhancement of the QFI. Figure\,\ref{fig3}(a) shows an enhancement with 7 orders of magnitude for the QFI, corresponding to a dramatic reduction of the optimal sensitivity $\Delta \mathbb{B}_z(\tau_1)=1/\sqrt{M\mathcal{F}_q{(\tau_1)}}$. As shown in Fig.\,\ref{fig3}(b), the optimal sensitivity $\Delta \mathbb{B}_z(\tau_1)$ can reach to the order of $10^{-15}-10^{-17}\rm{(T/\sqrt{Hz})}$ for a wide parameter region in terms of the mean photon numbers ${\mathcal{N}}_1$ and ${\mathcal{N}}_2$. Equation~(\ref{QFI}) also indicates that the resources ${\mathcal{N}}_1$ and ${\mathcal{N}}_2$ exert different influences on the ultimate lower bound of sensitivity\,\cite{SM}. We note that the QFI at $\tau_1$ does not depend on the actual value of $B_z$.

Next, let us calculate the CFI related to a specific measurement on the quadrature $X_{\theta}=(a_1e^{-i\theta}+a_1^{\dag}e^{i\theta})/\sqrt{2}$, where $\theta$ is the phase of local oscillator. At the first decoupled time $\tau_1$, the CFI is given by\,\cite{SM}
\begin{align}
\!\!\!\mathcal{F}_c(\tau_1)\!=\!\frac{32\pi^2m\lambda_1^2L^2\alpha_{\rm mag}^2}{\hbar\omega_mE^2}e^{12r}(\alpha_{\rm Re}\sin\theta\!-\!\alpha_{\rm Im}\cos\theta)^2.\!
\end{align}
Evidently, it consists precisely with the QFI shown in Eq.\,(\ref{QFI}) when $\theta=\pi/2$ ($\theta=0$) and $\alpha$ is a real (imaginary) number, which means that the momentum (position) measurement can saturate the optimal sensitivity in the absence of system dissipation.

\begin{figure}
	\includegraphics[width=8.4cm]{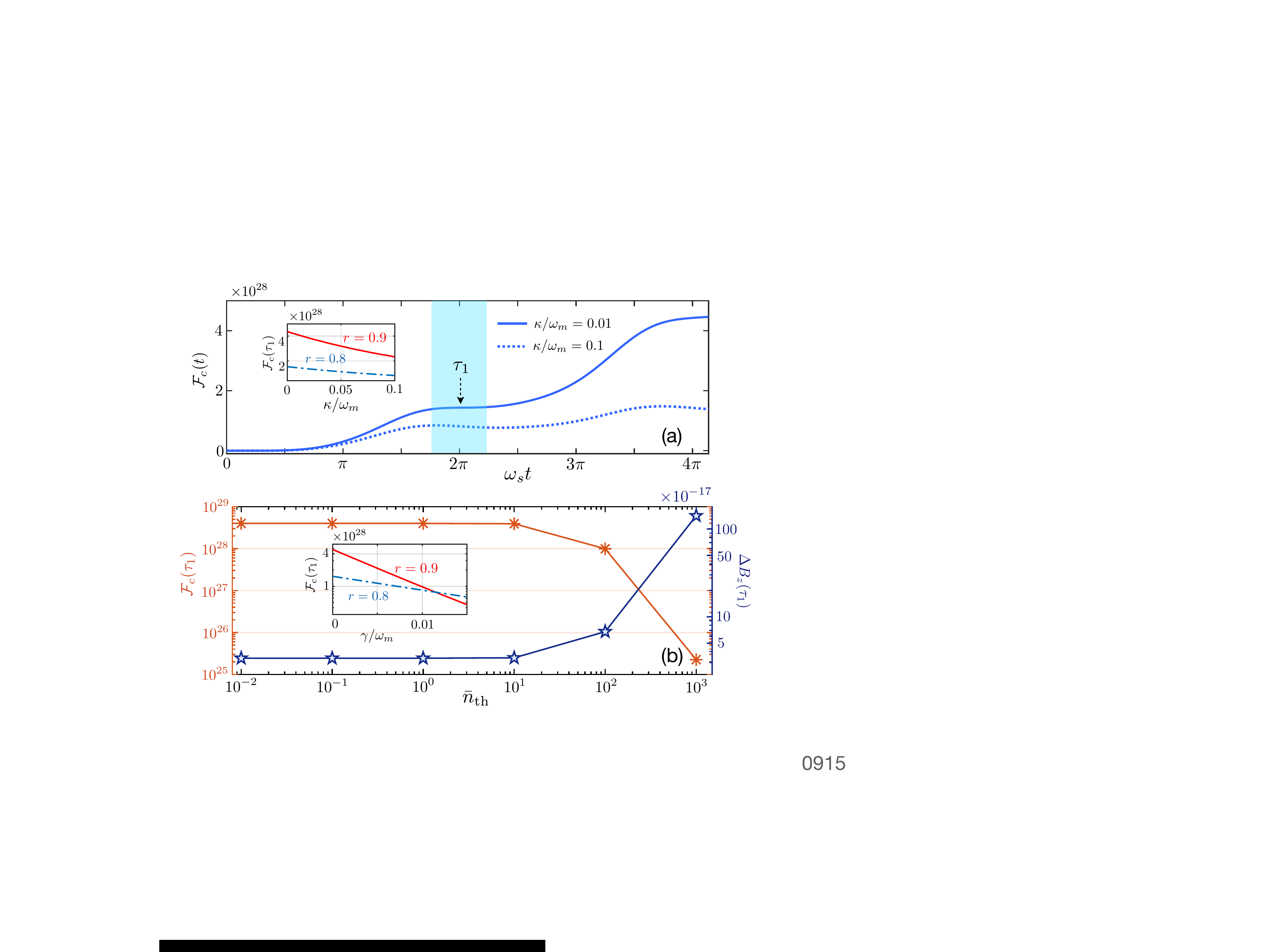}
	\caption{(a) Time evolution of $\mathcal{F}_{c}(t)$ for different values of $\kappa/\omega_m$ in the presence of mechanical dissipation $\gamma/\omega_m=0.01$, $\bar{n}_{\rm th}=10$ and considering $r=0.8$. The shaded area indicates the time-window of detection with nearly flat $\mathcal{F}_{c}(t)$ in the vicinity of the first decoupled time $\tau_1$. (b) CFI $\mathcal{F}_{c}(\tau_1)$(left y-axis) and specific sensitivity $\Delta B_z (\tau_1)$ in units of ${\rm T/\sqrt{Hz}}$ (right y-axis) versus mechanical thermal phonon number $\bar{n}_{\rm th}$ for $\kappa/\omega_m=0.01$ and $\gamma/\omega_m=0.001$. The insets show the influence of $\kappa$ and $\gamma$ on $\mathcal{F}_{c}(\tau_1)$ for different values of squeezing parameter $r$.}
	\label{fig4}
\end{figure}

In the practical situation, the dissipation caused by the system-bath coupling should be taken into account. Then the dynamics of system is dominated by the master equation
\begin{align}
\!\!\!\dot{\rho}\!=\!-\frac{i}{\hbar}[H,\rho]\!+\!\kappa\mathcal{D}[a_1]\rho\!+\!\gamma(\bar{n}_{\rm th}\!\!+\!1)\mathcal{D}[b]\rho\!+\!\gamma \bar{n}_{\rm th}\mathcal{D}[b^{\dag}]\rho\label{rho},
\end{align}
where $\kappa (\gamma)$ is the cavity (mechanical) decay rate, $\bar{n}_{\rm th}$ is the thermal phonon number of the mechanical mode, and $\mathcal{D}[o]\rho=o\rho o^{\dag}-(o^{\dag}o\rho+\rho o^{\dag}o)/2$. Here we have considered the cavity mode $a_2$ being in the coherent state $|\xi\rangle$ for Hamiltonian $H$. Performing a momentum homodyne measurement (i.e., $\theta=\pi/2$) on cavity $a_1$, in Fig.\,\ref{fig3}(c) and Fig.\,\ref{fig4}, we numerically demonstrate the influence of system dissipation on the sensitivity limit, i.e., $\Delta B_z(\tau_1)=1/\sqrt{M\mathcal{F}_c{(\tau_1)}}$, of magnetometer\,\cite{johansson201201}.

With the practical experimental parameters, Figure\,\ref{fig3}(c) shows that the specific sensitivity $\Delta B_z(\tau_1)$ obtained in the presence of system decay and noise, still can fit well with the optimal one $\Delta \mathbb{B}_z(\tau_1)$ from the QFI without system dissipation. This consistency is only broken weakly when one increases the squeezing parameter $r$ to a large value. This can be explained as follows. On the one hand, system dissipation has little effect on the CFI $\mathcal{F}_{c}(\tau_1)$ in the case of weak squeezing parameter $r$ [see the inserts of Fig.\,\ref{fig4}]. More importantly, the periodic optomechanical decoupling makes the CFI robust against the mechanical thermal noise. As shown in Fig.\,\ref{fig4}(b), a high sensitivity reaching to the order of $\sim 10^{-15}{\rm T/\sqrt{Hz}}$ is allowed even when $\bar{n}_{\rm th}=10^3$. This means that the mechanical ground state cooling is not necessary for obtaining high-precision magnetometer in our proposal. On the other hand, the strong mechanical squeezing amplifies the effect of mechanical dissipation on CFI via effectively heating the environment (see the insert of Fig.\,\ref{fig4}(b) and Fig.\,S5 in supplementary material\,\cite{SM}), which leads to the weak disagreement between the measurement-specific sensitivity limit and the fundamental bound of sensitivity in the case of large values of $r$.

\emph{Discussion of experimental feasibility.---}
Regarding experimental implementations, while we have considered here a Febry-P\'{e}rot cavity with the membrane-in-the-middle configuration, our versatile proposal is not limited to this particular architecture. Based on the excellent controllability of the SQUID\,\cite{You2011,Xiang2013}, the transmission-line (TL) resonator coupled to a SQUID-terminated TL resonator is a promising platform to realize dual-coupling optomechanical system\,\cite{Johansson2014,Kim2015}. Recently, the dual-coupling optomechanics was also demonstrated in photonic crystal cavities\,\cite{Kalaee2016,Brunelli2018} and whispering gallery microcavities\,\cite{Hao-Kun2012}. Based on recent optomechanical experiments\,\cite{thompson2008,Sankey2010,Ockeloen-Korppi2018}, here the system parameters can be chosen as $m=4\times10^{-8}$ g, $\omega_m=2\pi\times 134$ kHz, $\lambda_1=8.4$ kHz, $\lambda_2=0.08$ Hz, $\kappa = 8.4$ kHz, $\gamma = 840 $ Hz, and $\mathcal{N}_1=10^6$. Then, with a cycle time on tens of $\mu$s, our work theoretically predicts that the sensitivity in the range of $10^{-15}\!-\!10^{-17}\,{\rm T/\sqrt{Hz}}$ can be realized with the achievable squeezing parameter $r\in[0,0.9]$. Note that most of the above results are obtained in the case of performing the homodyne measurement at the first decoupled time $\tau_1$. Fortunately, our proposal is robust against the detection time, i.e., the CFI in the vicinity of $\tau_1$ is nearly flat as shown in Figure\,\ref{fig4}(a). In other words, our proposal exhibits a wide time-window to perform measurement with sensitivity reaching to the order of $10^{-17}{\rm T}/\sqrt{\rm Hz}$.

Moreover, the experimental implementation of our proposal relies on the thin membrane held by Terfenol-D rods, which gives an effective magnetic potential on the Hamiltonian. The conversion efficiency from the applied magnetic field to an effective force on the mechanical oscillator, is determined by the magnetic actuation constant $c_{\rm act}$. The precise experimental determination of the magnetic actuation constant is of key importance for measuring afterwards accurate values of the magnetic field. This magnetic constant depends heavily on the specific geometrical configuration of the rods with respect to the membrane. To optimize the magnetostrictive effect, the applied magnetic field needs to be paralleled to the magnetostricitive direction of the Terfenol-D rods. In addition, here we considered the Terfenol-D rods of  the length $\sim 10^{-5}$ m, thus it is approximatively valid to consider the magnetostrictive material to be in the whole homogeneous magnetic field, which also allows the system to have a large actuation constant.

Despite here we focus on detecting a static magnetic field, our proposal, in principle, can also be applied to probe the alternating magnetic fields. The corresponding frequency response characteristics are discussed in Sec.\,VII of the supplementary material\,\cite{SM}. Referring to the parameters used in optomechanical experiments \,\cite{Bei-Bei2021,thompson2008}, we numerically simulate the displacement noise power spectrum and the corresponding force sensitivity at different probe powers (see Fig.\,S6 in the supplementary material). We find that the thermal-noise-limited frequency range covers $0\sim 380$ kHz with the central resonant frequency 250 kHz in the case of the probe power at $20 \,{\rm nW}$. Therefore, similar as a general optomechanical system\,\cite{Bei-Bei2021}, here the dual-coupling magnetometer also has a broad bandwidth, when it is used as a resonant sensor.

\emph{Conclusions.--} We have presented a protocol to measure the weak magnetic field using dual-coupling optomechanics. The sensitivity could be enhanced to the order of $10^{-15}-10^{-17}\,{\rm T/\sqrt{Hz}}$ by adjusting the photon numbers of two cavity modes. This enhancement originally comes from the periodic mechanical squeezing, which greatly amplifies both the signal to be measured and the transducing rate of signal. We stress out that this periodic squeezing effect is self-sustained, which avoids the complicated process of preparing squeezed states. Our proposal, with wide time-window of detection, is robust against the mechanical thermal noise, and hence the ground state cooling of mechanical mode is not necessary. This work might inspire the studies of high-precision measurements of various physical quantities based on dual-coupling optomechanical systems.

{\it Acknowledgments}.--We thank Bei-Bei Li for fruitful discussions and valuable comments. This work is supported by the National Key Research and Development Program of China grant 2021YFA1400700 and the National Science Foundation of China (Grant Nos. 11822502, 11974125, 11875029, 12175075 and 11805073).

%\bibliography{ref}

%

\newpage

%%%%%%%%%% Merge with supplemental materials %%%%%%%%%%
\onecolumngrid

%%%%%%%%%% Prefix a "S" to all equations, figures, tables and reset the counter %%%%%%%%%%
\setcounter{equation}{0} \setcounter{figure}{0}
\setcounter{table}{0}
\setcounter{page}{1}\setcounter{secnumdepth}{3} \makeatletter
\renewcommand{\theequation}{S\arabic{equation}}
\renewcommand{\thefigure}{S\arabic{figure}}
\renewcommand{\bibnumfmt}[1]{[S#1]}
\renewcommand{\citenumfont}[1]{S#1}
\renewcommand\thesection{S\arabic{section}}
%%%%%%%%%% Prefix a "S" to all equations, figures, tables and reset the counter %%%%%%%%%%

\begin{center}
{\large \bf Supplementary Material for\\ ``Quantum Magnetometer with Dual-Coupling Optomechanics''}
\end{center}

\begin{center}
Gui-Lei Zhu$^{1}$, Jing Liu$^{1,2}$, Ying Wu$^1$,
and Xin-You L\"{u}$^{1,^*}$
\end{center}

\begin{minipage}[]{18cm}
\small{\it
\centering $^{1}$School of Physics, Huazhong University of Science and Technology, Wuhan 430074, China \\
\centering $^{2}$MOE Key Laboratory of Fundamental Physical Quantities Measurement and PGMF, \\
\centering Huazhong University of Science and Technology, Wuhan 430074, China \\}

\end{minipage}

\vspace{8mm}
\tableofcontents
\section*{Overview of the Supplemental Material}
In this Supplementary Material, we present the technical details of the dual-coupling optomechanical magnetometer considered in the main text. In Sec.\,\ref{section1}, we list the main symbols and parameters used in this work. In Sec.\,\ref{section2}, we give a detailed description of the mechanical responses of the magnetostrictive material exposed to external magnetic fields, and present the system Hamiltonian in the form of second quantization. We also derive the effective Hamiltonian in the squeezed frame when the ancillary cavity mode $a_2$ is in the coherent state. In Sec.\,\ref{section3}, we present the detailed derivations of system evolution as well as the periodic squeezing effect during the dynamical evolution. The detailed derivation of state tomography at the first decoupled time $\tau_1$ is presented in Sec.\,\ref{section4}. Moreover, in Sec.\,\ref{section5}, we derive the QFI and the fundamental bound to the sensitivity of the proposed magnetometer. In Sec.\,\ref{section6},  considering a momentum measurement on the cavity mode $a_1$, we derive the CFI corresponding to the specific sensitivity limit of the proposed magnetometer. By numerically solving the master equation, we illustrate the influence of system dissipations on the CFI and the associated specific sensitivity. {In Sec.\,\ref{section7}, we discuss the frequency response characteristics (e.g., the bandwidth), when the proposed dual-coupling optomechanical magnetometer is used to detect the alternating magnetic fields as a resonant sensor.}

\section{System Parameters}\label{section1}
In Table \ref{tab:tableS1}, we list the main parameters used in our proposal. All of them are experimentally feasible in the state-of-the-art setups.
\begin{table}[h]
\caption{\label{tab:tableS1} Main symbols and parameters have been used in this work. If not specified, these parameters are used in all calculations.}
{\begin{ruledtabular}
	\begin{tabular}{lll}
	\textbf{Symbols}&\textbf{Parameters}&\textbf{Value}\\
	\colrule
	$m$ & Effective mass of mechanical oscillator&$4\times 10^{-8}{\rm g}$\,\cite{Thompson2008} \\
	$\omega_m$&Mechanical frequency&{$2\pi\times 134\,{\rm kHz}$}\,\cite{Thompson2008}\\
	$\lambda_1$ &Radiation-pressure coupling strength &{$ 8.4\,{\rm kHz}$}\\
	$\lambda_2$&Quadratic optomechanical coupling strength&$ 0.08{\rm Hz}$\\
	$\gamma$&Mechanical decay rate&$(2.5\sim 12)\,{\rm kHz}$\\
	$\kappa$&Cavity decay rate&{$ (8.4\sim 84)\,{\rm kHz}$}\\
	$\bar{n}_{\rm th}$& Thermal phonon occupation &$10^{-2}\sim 10^3$\\
	$k=m\omega_m^2$& Spring constant&$28\,{\rm N m^{-1}}$\,\cite{Thompson2008} \\
	$L$&The length of Terfenol-D rods&630\,$\mu {\rm m}$\\
	$\alpha_{\rm mag}$&	Magnetostrictive coefficient along $z$ axis&$5\times 10^8\, {\rm N T^{-1}m^{-2}}$\cite{Verhoeven1990}\\
	$E$ &Young's modulus of Terfenol-D& $30\,{\rm GPa}$\\
	$c_{\rm act}$&Magnetic actuation constant& $2.98\times 10^{-4}\,{\rm NT^{-1}}$\\
	$\mathcal{N}_1=\langle a_1^{\dag}a_1\rangle$&Mean photon number of cavity mode $a_1$&$10^6$
	\\
	$\mathcal{N}_2=\langle a_2^{\dag}a_2\rangle$&Mean photon number of cavity mode $a_2$ &$0\sim2.45\times10^6$\\
	$r=-0.25{\rm ln}(1-4\lambda_2\mathcal{N}_2/\omega_m)$ &Squeezing parameter &$ 0\sim 0.9$
	\end{tabular}
	\end{ruledtabular}}\label{tableS1}
	\end{table}

\section{System Hamiltonian}\label{section2}
The quantum system considered in this work is a dual-coupling optomechanical system, where two optical modes simultaneously coupled to the same mechanical oscillator, with the Hamiltonian
\begin{align}	\label{Hin}
H=\hbar \omega_1a_1^{\dag}a_1+\hbar\omega_2 a_2^{\dag}a_2+\hbar\omega_mb^{\dag}b-\hbar\lambda_1a_1^{\dag}a_1(b^{\dag}+b)-\hbar\lambda_2a_2^{\dag}a_2(b^{\dag}+b)^2+B_zc_{\rm act}z,
\end{align}
where $a_1$, $a_2$ and $b$ are the annihilation operators of optical cavity modes (with frequency $\omega_1, \omega_2$) and the mechanical oscillator (with frequency $\omega_m$). Here $\lambda_1$ and $\lambda_2$ are the radiation-pressure coupling and quadratic optomechanical coupling strengths, respectively. The magnetostrictive potential has been included in the last term on the right-hand side. Let us first discuss in detail the generation of magnetostrictive potential.
	\begin{figure}[h]
	\includegraphics[width=15.0cm]{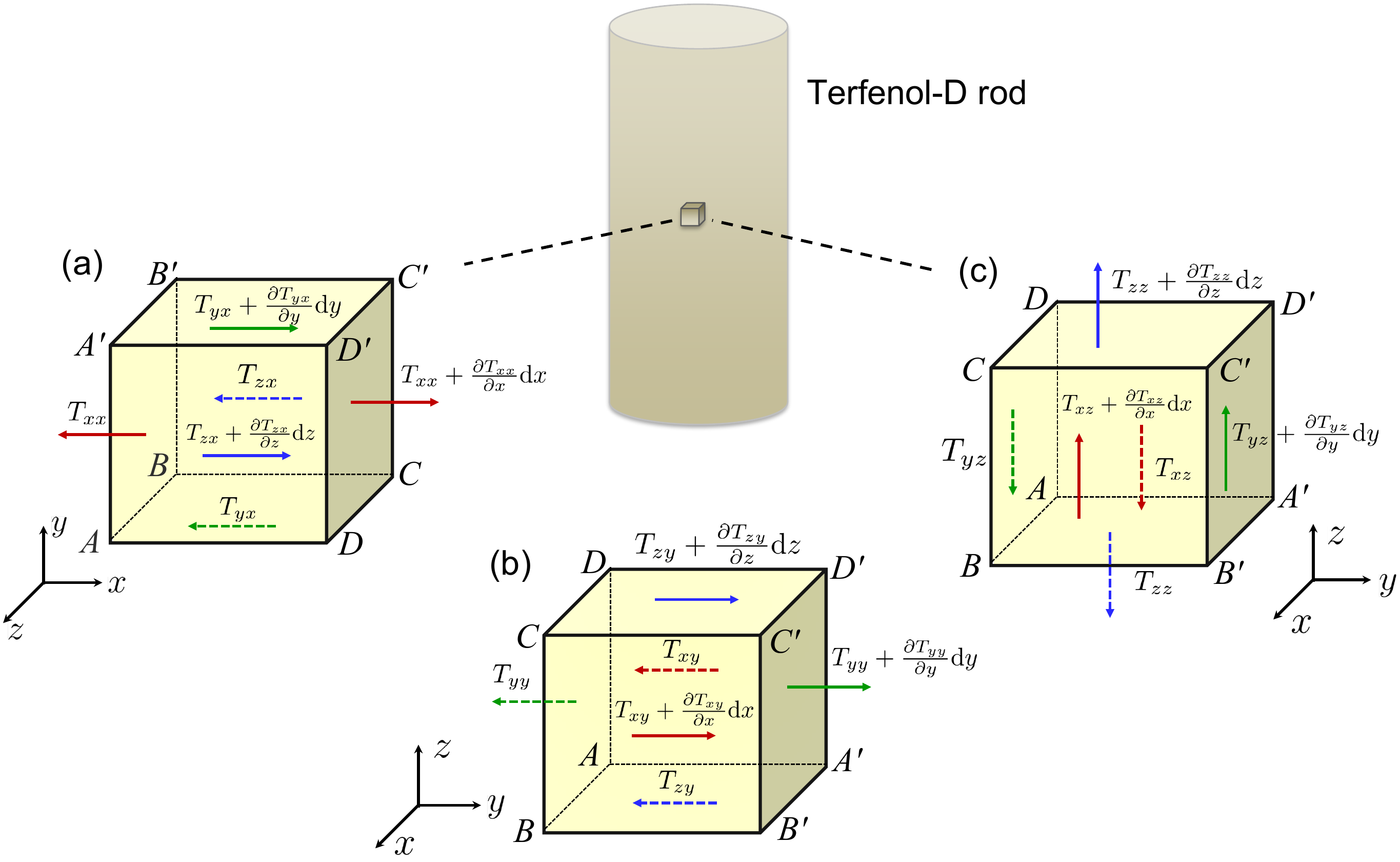}
	\caption{Infinitesimal volume element taken from interior of Terfenol-D rod. (a-c) show the variation of stress due to the presence of body force density $
		\emph{\textbf f}_{\rm mag}$ along $x, y, z$ axes, respectively. Stress tensor elements $T_{ij} (i=j)$ and $T_{ij} (i\ne j)$ represent the pure stress and the shear stress, respectively.}
	\label{figS1}
\end{figure}

\subsection{Generation of magnetostrictive potential}
Schematic setup of our proposal is shown in Fig.\,1 of the main text. We start by describing the connection between strain and the magnetic force acting on Terfenol-D rods. The Terfenol-D rods are subjected to a magnetic field with body force density $\emph{\textbf{f}}_{\rm mag}=f_x\vec{x}+f_y\vec{y}+f_z\vec{z}$, where $\vec{x}, \vec{y}$ and $\vec{z}$ are unit vectors along $x, y$ and $z$ axes, respectively. We divide up this object into infinitesimally small cubic volume element, with its edges aligned with the coordinate axes, as shown in Fig.\,\ref{figS1}. The small volume element has the size of ${\rm d}x{\rm d}y{\rm d}z$. Essentially, the cube is acted upon the forces from the surrounding solid, and the magnetostrictive stress will vary with position. Therefore, we expand the stress tensor $\textbf T$ in a Taylor series as a function of position within the cube. In a static equilibrium condition, the forces along $x, y, z$ coordinate axes are balanced [see Fig.\,{\ref{figS1}], which is given by\,\cite{Cleland2003}
\begin{align}
		&(T_{xx}\!+\!\frac{\partial T_{xx}}{\partial{x}}{\rm d}x){\rm d}y{\rm d}z\!-\!T_{xx}{\rm d}y{\rm d}z\!+\!(T_{yx}\!+\!\frac{\partial T_{yx}}{\partial{y}}{\rm d}y){\rm d}x{\rm d}z\!-\!T_{yx}{\rm d}x{\rm d}z\!+\!(T_{zx}\!+\!\frac{\partial T_{zx}}{\partial{z}}{\rm d}z){\rm d}x{\rm d}y\!-\!T_{zx}{\rm d}x{\rm d}y\!+\!f_x{\rm d}x{\rm d}y{\rm d}z\!=\!0,\\
		&(T_{xy}\!+\!\frac{\partial T_{xy}}{\partial{x}}{\rm d}x){\rm d}y{\rm d}z\!-\!T_{xy}{\rm d}y{\rm d}z\!+\!(T_{yy}\!+\!\frac{\partial T_{yy}}{\partial{y}}{\rm d}y){\rm d}x{\rm d}z\!-\!T_{yy}{\rm d}x{\rm d}z\!+\!(T_{zy}\!+\!\frac{\partial T_{zy}}{\partial{z}}{\rm d}z){\rm d}x{\rm d}y\!-\!T_{zy}{\rm d}x{\rm d}y\!+\!f_y{\rm d}x{\rm d}y{\rm d}z\!=\!0,\\
		&(T_{xz}\!+\!\frac{\partial T_{xz}}{\partial{x}}{\rm d}x){\rm d}y{\rm d}z\!-\!T_{xz}{\rm d}y{\rm d}z\!+\!(T_{yz}\!+\!\frac{\partial T_{yz}}{\partial{y}}{\rm d}y){\rm d}x{\rm d}z\!-\!T_{yz}{\rm d}x{\rm d}z\!+\!(T_{zz}\!+\!\frac{\partial T_{zz}}{\partial{z}}{\rm d}z){\rm d}x{\rm d}y\!-\!T_{zz}{\rm d}x{\rm d}y\!+\!f_z{\rm d}x{\rm d}y{\rm d}z\!=\!0.
		\end{align}
When ${\rm d}x{\rm d}y{\rm d}z\ne 0$, the above equations can be simplified as
		\begin{align}\label{EQ5}
		\frac{\partial T_{xx}}{\partial{x}}+\frac{\partial T_{yx}}{\partial{y}}+\frac{\partial T_{zx}}{\partial{z}}+f_x=0,\\\label{EQ6}
		\frac{\partial T_{xy}}{\partial{x}}+\frac{\partial T_{yy}}{\partial{y}}+\frac{\partial T_{zy}}{\partial{z}}+f_y=0,\\\label{EQ7}
		\frac{\partial T_{xz}}{\partial{x}}+\frac{\partial T_{yz}}{\partial{y}}+\frac{\partial T_{zz}}{\partial{z}}+f_z=0.
		\end{align}
Next, we assume a homogeneous magnetic field $B_z$ oriented in the $z$ direction, and neglect the shear strain in the direction perpendicular to the tension. Then the magnetostrictive material only stretch along $z$ direction with magnetostrictive coefficient $\alpha_{\rm mag}$, and the magnetostrictive-induced stress tensor $\textbf T$ has only a single component $T_{zz}=\alpha_{\rm mag}B_z$. Simplifying Eqs.\,(\ref{EQ5}-\ref{EQ7}), we obtain the body force density
\begin{align}
f_{z}=-\frac{\partial T_{zz}}{\partial z}.
\end{align}
Under the action of the magnetic field, the magnetostrictive actuators stretch and further move the mount of suspended membrane (see Fig.\,1 shown in main text). The displacement of an infinitesimally small cubic volume element at initial position $\emph{\textbf r}$ and time $t$ is~\cite{Briant2003}
\begin{align}
{\emph{\textbf u}}(\emph{\textbf r},t)={\Psi}_q(\emph{\textbf r})X_q(t),
\end{align}
where ${\Psi}_q(\emph{\textbf r})$ is the position-dependent mode shape function of eigenmode $q$, which is normalized by $\int_{V}{\Psi}_p(\emph{\textbf r})\cdot{\Psi}_q(\emph{\textbf r}){\rm d}x{\rm d}y{\rm dz}=\delta_{pq}V$ with $V$ being the spatial volume of oscillator. In addition, $X_q(t)$ depends on the force applied on the membrane. Here, we consider the membrane having a single mechanical eigenmode with frequency $\omega_m$ and effective mass $m$. The driving force received by the mechanical oscillator is~\cite{fforstner201201}
\begin{align}
F_{\rm mag}=\int \emph{\textbf f}_{\rm mag}(\emph{\textbf r})\cdot \emph{\textbf u}(\emph{\textbf r}){\rm d}V=\int f_z{\Psi}_z(\emph{\textbf r}){\rm d}V=B_zc_{\rm act}.
\end{align}
where \begin{align}\label{cact}
c_{\rm act}=m\omega_m^2L\frac{\alpha_{\rm mag}}{E},\end{align}
is the magnetic actuation constant charactering how well the magnetic field is converted into an applied force on the oscillator. Here $L$ is the length of Terfenol-D rod and $E$ is the Young's modulus of Terfenol-D. The potential caused by the magnetic field can be written as
\begin{align}
\Delta U=B_zc_{\rm act}z,
\end{align}
where $z=\sqrt{\hbar/2m\omega_m}(b^{\dag}+b)$ is the mechanical position operator. So far, we have obtained the magnetostrictive potential.

\subsection{Hamiltonian in the squeezed frame}
By considering the ancillary mode $a_2$ in a coherent state $|\xi\rangle$, the system Hamiltonian can be reduced to
\begin{align}\label{EQ:H}
H=\hbar\omega_1a_1^{\dag}a_1+\hbar\omega_2|\xi|^2+\hbar \omega_mb^{\dag}b-\hbar\lambda_1a_1^{\dag}a_1(b^{\dag}+b)-\hbar\lambda_2|\xi|^2(b^{\dag}+b)^2+B_zc_{\rm act}z.
\end{align}
In deriving Hamiltonian (\ref{EQ:H}), we have assumed $\langle \xi|a_2^{\dag}a_2|\xi\rangle\approx |\xi|^2 $. Even though the coherent state $|\xi\rangle$ is not the eigenstate of photon number operator $a_2^{\dag}a_2$, the relative fluctuation is
\begin{align}
F_{\rm re}&=\frac{\sqrt{\langle \xi|a_2^{\dag}a_2a_2^{\dag}a_2|\xi\rangle-\langle\xi|a_2^{\dag}a_2|\xi\rangle^2}}{|\xi|^2}=\frac{1}{|\xi|}.
\end{align}
It is clearly shown that the higher $\xi$, the smaller relative fluctuation. In our calculations, we considered $\xi\approx1550$}, whose fluctuation is much smaller than mean-photon number and can be neglected safely.	
\begin{figure}[h]
	\includegraphics[width=12.0cm]{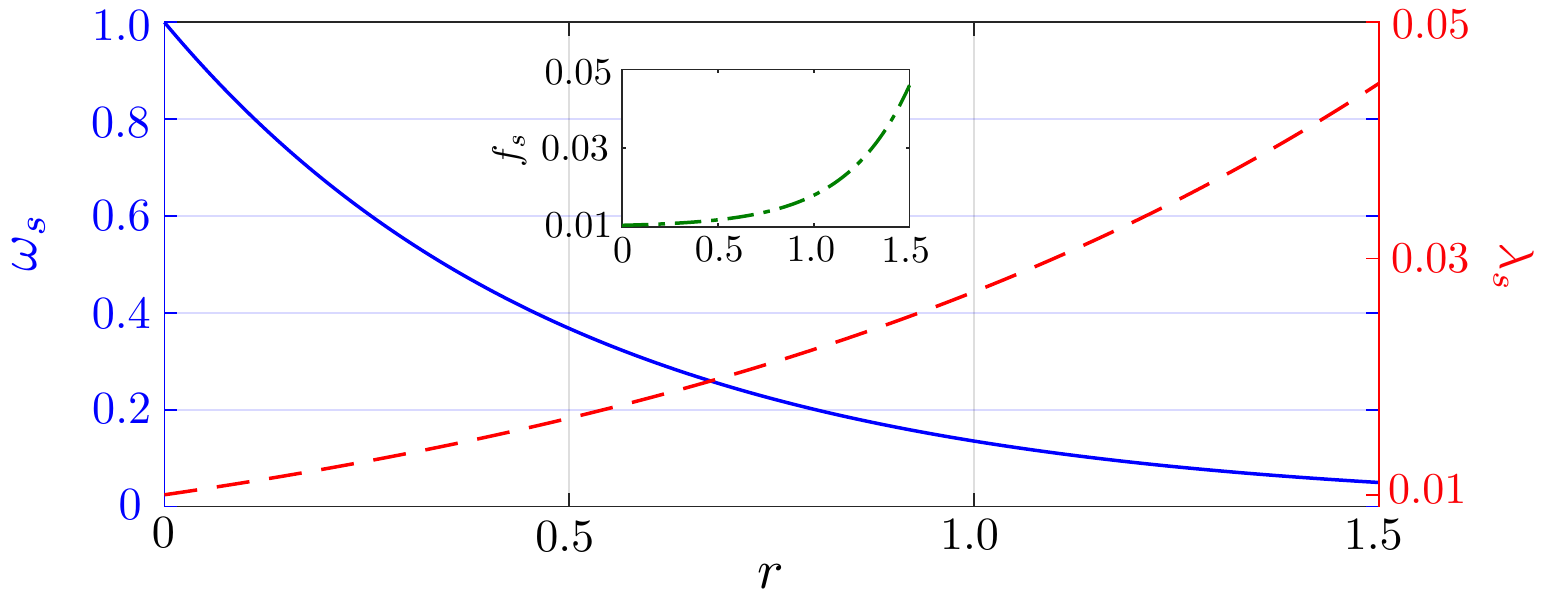}
	\caption{Effective mechanical frequency $\omega_s$ (left-axis), optomechanical coupling strength ${\lambda}_s$(right-axis) and $f_s$(inset)  versus the squeezing parameter $r$ with $\omega_m=1$. }
	\label{figS2}
\end{figure}
Next, we apply a Schrieffer-Wolff transformation $H_s=S(r)HS^{\dag}(r)$ to Eq.\,(\ref{EQ:H}) with squeezing operator $S(r)=\exp{[r(b^2-b^{\dag 2})/2]}$  and  squeezing parameter $r=-(1/4)\ln(1-4\lambda_2|\xi|^2/\omega_m)$. In view of
\begin{align}
S(r)bS^{\dag}(r)=b\cosh r+b^{\dag}\sinh r,\\
S(r)b^{\dag}S^{\dag}(r)=b^{\dag}\cosh r+b\sinh r,
\end{align}
we obtain the Hamiltonian in the squeezed frame is
	\begin{align}\label{EQ:Heff}
	H_{s}/\hbar=\omega_1a_1^{\dag}a_1+\omega_sb^{\dag}b-\lambda_sa_1^{\dag}a_1(b^{\dag}+b)+f_s(b^{\dag}+b)+E_0.
	\end{align}
where $\omega_s=\omega_me^{-2r}$, $\lambda_s=\lambda_1e^r$ and $f_s=fe^r=\sqrt{{1}/{(2m\hbar \omega_m)}}B_zc_{\rm act}e^r$. Here $E_0=\omega_m(e^{-2r}-1)/2+\omega_2|\xi|^2$ is the constant term and will be omitted in the following discussions for the sake of simplicity. In Fig.\,\ref{figS2} we plot the effective mechanical frequency $\omega_s$ and optomechanical coupling strength $\lambda_s$ versus the squeezing parameter $r$. Evidently, increasing the parameter $r$ enables an exponential reduction of mechanical frequency, but an exponential enhancement of the radiation-pressure optomechanical coupling. Moreover, from the fourth term of Hamiltonian $H_{s}$ we observe that the magnetic signal to be measured is {\it exponentially amplified} [also see the insert of Fig.\,\ref{figS2}], which directly gives rise to the enhancement of estimation precision of magnetic field.

\section{System dynamics with periodic mechanical squeezing}\label{section3}
\subsection{Periodic opto-mechanical decoupling}
 Starting from the original Hamiltonian Eq.\,(\ref{EQ:H}) under the condition of applying the mean field approximation for mode $a_2$, we can write the evolution operator of system as%\,\cite{Armata2017}
 \begin{align}\label{EQ:Ut}
 U(t)&=e^{i(\tilde{\lambda}_sa_1^{\dag}a_1-\tilde{f}_s)^2[\omega_st-\sin(\omega_st)]}S^{\dag}(r)e^{(\tilde{\lambda}_sa_1^{\dag}a_1-\tilde{f}_s)(\mu b^{\dag}-\mu^*b)}e^{-ib^{\dag}b\omega_st}S(r),\nonumber\\
 &=e^{i(\tilde{\lambda}_sa_1^{\dag}a_1-\tilde{f}_s)^2[\omega_st-\sin(\omega_st)]}S^{\dag}(r)S(r')e^{(\tilde{\lambda}_sa_1^{\dag}a_1-\tilde{f}_s)(\bar{\mu} b^{\dag}-\bar{\mu}^*b)}e^{-ib^{\dag}b\omega_st},
 \end{align}
where
\begin{align}
\tilde{\lambda}_s&=\lambda_s/\omega_s=\frac{\lambda_1}{\omega_m}e^{3r},\\\tilde{f}_s&=f_s/\omega_s=B_zc_{\rm act}\sqrt{\frac{1}{2m\hbar \omega_m^3}}e^{3r},\end{align}
are the rescaled parameter, and $\bar{\mu}=(1-e^{-i\omega_st})(\cosh r-e^{-i\omega_st} \sinh r)$,
$S(r')=\exp{[(r'^*b^2-r'b^{\dag 2})/2]}$ with $r'=re^{-2i\omega_st}$.  In derivating Eq.\,(\ref{EQ:Ut}), we have adopted a rotating frame with $\exp(-i\tilde{\omega}_1a_1^{\dag}a_1)$
where $\tilde{\omega}_1=\omega_1/\omega_s$. Under the condition that the optical and mechanical modes are initially in the coherent state $|\alpha\rangle|\beta\rangle$ with $\beta=\beta_{\rm Re}+i\beta_{\rm Im}$, the state at time $t$ is
	\begin{align}
	|\Psi(t)\rangle&=e^{i(\tilde{\lambda}_sa_1^{\dag}a_1-\tilde{f}_s)^2[\omega_st-\sin(\omega_st)]}S^{\dag}(r)S(r')e^{(\tilde{\lambda}_sa_1^{\dag}a_1-\tilde{f}_s)(\bar{\mu} b^{\dag}-\bar{\mu}^*b)}e^{-ib^{\dag}b\omega_st}|\alpha\rangle|\beta\rangle\nonumber\\
	&=e^{-|\alpha|^2/2}\sum_{n=0}^{\infty}\frac{\alpha^n}{\sqrt{n!}}e^{i(\tilde{\lambda}_sn-\tilde{f}_s)^2[\omega_st-\sin (\omega_st)]}|n\rangle\left(S^{\dag}(r)S(r')e^{(\tilde{\lambda}_sn-\tilde{f}_s)(\bar{\mu}b^{\dag}-\bar{\mu}^*b)}|e^{-i\omega_st}\beta\rangle\right)\nonumber\\
	&=e^{-|\alpha|^2/2}\sum_{n=0}^{\infty}\frac{\alpha^n}{\sqrt{n!}}e^{i(\tilde{\lambda}_sn-\tilde{f}_s)^2[\omega_st-\sin (\omega_st)]}|n\rangle\left(S^{\dag}(r)S(r')D[(\tilde{\lambda}_sn-\tilde{f}_s)\bar{\mu}]D(e^{-i\omega_st}\beta)|0\rangle\right) \nonumber\\
	&=e^{-|\alpha|^2/2}\sum_{n=0}^{\infty}\frac{\alpha^n}{\sqrt{n!}}e^{i(\tilde{\lambda}_sn-\tilde{f}_s)^2[\omega_st-\sin (\omega_st)]}e^{i(\tilde{\lambda}_sn-\tilde{f}_s)e^{-r}[\beta_{\rm Re}\sin\omega_st-\beta_{\rm Im}(\cos \omega_st-1)e^{2r}]}|n\rangle\left(S^{\dag}(r)S(r')|\varphi_n(t)\right\rangle),
	\label{EQ:Psit}
	\end{align}
	where $\varphi_n(t)=e^{-i\omega_st}\beta+(\tilde{\lambda}_sn-\tilde{f}_s)\bar{\mu}$. Here $D(\beta)=e^{-|\beta|^2/2}e^{\beta b^{\dag}}e^{-\beta^*b}$ is the displacement operator, and we have used $D(\alpha+\beta)=D(\alpha)D(\beta)\exp[-i{\rm Im} (\alpha\beta^*)]$. By tracing out the mechanical part, we obtain the reduced cavity state
	\begin{align}
	\rho_c(t)&=e^{-|\alpha|^2}\sum_{n,n'}\left[\frac{\alpha^n(\alpha^*)^{n'}}{\sqrt{n!n'!}}e^{i\left[(\tilde{\lambda}_sn-\tilde{f}_s)^2-(\tilde{\lambda}_sn'-\tilde{f}_s)^2\right][\omega_st-\sin(\omega_st)]}e^{i\tilde{\lambda}_s(n-n')e^{-r}[\beta_{\rm Re}\sin \omega_st-\beta_{\rm Im}(\cos \omega_st-1)e^{2r}]}\right.\nonumber\\
	&\left.
	\times e^{-(|\varphi_n|^2+|\varphi_{n'}|^2)/2+\varphi_{n'}^*\varphi_n}|n\rangle\langle n'|\right].
	\end{align}
Interestingly, at time $\tau_m=2m\pi /\omega_s (m=1,2,3...)$, the parameters $\bar{\mu}=0, r'=r, S^{\dag}(r')S(r)=1$ and $|\varphi_n(\tau_m)\rangle=|\beta\rangle$, and then Eq.\,(\ref{EQ:Psit}) is reduced to
	\begin{align}\label{EQ:Psitau}
	|\Psi(\tau_m)\rangle&=e^{-|\alpha|^2/2}\sum_{n=0}^{\infty}\frac{\alpha^n}{\sqrt{n!}}e^{i2m\pi(\tilde{\lambda}_sn-\tilde{f}_s)^2}|n\rangle|\beta\rangle,\nonumber\\
	&=e^{-|\alpha|^2/2}\sum_{n=0}^{\infty}\frac{\alpha^n}{\sqrt{n!}}e^{i\Phi_n(\tau_m)}|n\rangle|\beta\rangle.
	\end{align}
By tracing out the mechanical mode, the corresponding reduced density matrix becomes
	\begin{align}
	\rho_c(\tau_m)=e^{-|\alpha|^2}\sum_{n,n'}\frac{\alpha^n\alpha^{*n'}}{\sqrt{n!n'!}}e^{i[\Phi_n(\tau_m)-\Phi_{n'}(\tau_m)]}|n\rangle\langle n'|.
	\end{align}
It is shown from Eq.\,(\ref{EQ:Psitau}) that, at time $\tau_m$, the state of mechanical part back to its initial state (i.e., $|\beta\rangle$). This demonstrates that the mechanical and optical modes completely decoupled at this time. Meanwhile the signal to be measured has been transferred into the phase $\Phi_n(\tau_m)=2m\pi (\tilde{\lambda}_sn-\tilde{f}_s)^2$. The period of optical-mechanical decoupling is $\tau_1=2\pi/\omega_s$.
		
\subsection{Periodic mechanical squeezing effect}
\begin{figure}
	\includegraphics[width=12.0cm]{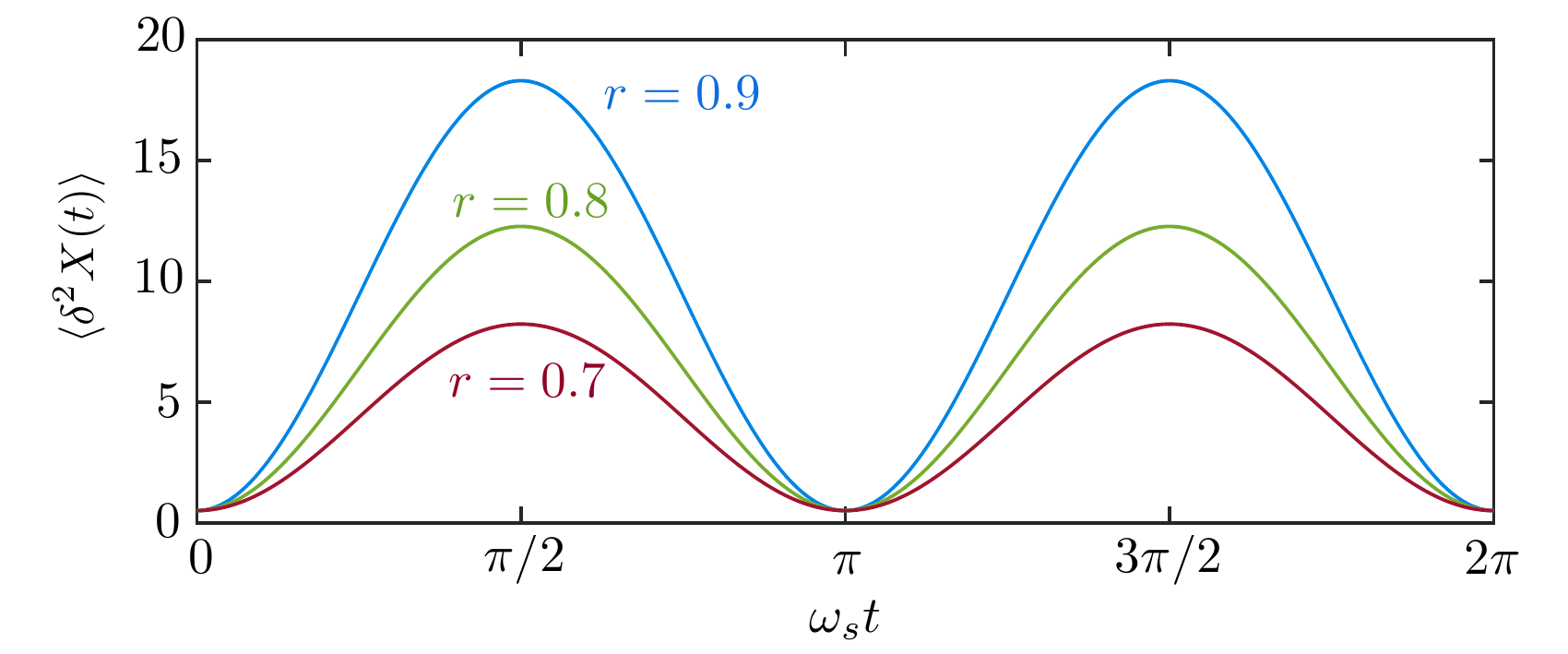}
	\caption{Variance of $X$ versus time for different squeezing parameter $r$.}
	\label{figS3}
\end{figure}

As shown in the main text, there is a periodic squeezing effect on the mechanical mode associated with the system evolution, which is also demonstrated in Eq.\,(\ref{EQ:Psit}). To qualitatively illustrate this periodic squeezing effect, for simplicity, we consider cavity $a_1$ in vacuum state and the auxiliary cavity $a_2$ with amplitude $\xi=1$, and neglect the magnetostrictive potential for a while. Then system Hamiltonian\,(\ref{EQ:H}) can be simplified to
\begin{align}\label{Hmod}
H/\hbar =\omega_m b^{\dagger }b-\lambda_2 \left( b^{\dagger
}+b\right) ^{2}. \end{align}
 Applying a squeezing transformation $H_{s}=S(r)HS^{\dag}(r) $ with $r=-0.25{\rm ln}(1-4\lambda_2/\omega_m)$,
 the resulting Hamiltonian\,(\ref{Hmod}) in squeezed frame is then of the form
 \begin{align}
 H_s/\hbar=\omega _{s}b^{\dagger }b, \end{align}
 where $ \omega _{s} =e^{-2r}\omega$.
Accordingly, the time evolution operator reads
 \begin{eqnarray*}
  U(t) =S^{\dagger }\left( r\right) e^{-i\omega _{s}tb^{\dagger }b}S\left(
  r\right),
 \end{eqnarray*}%
where $S(r) =e^{\left( 1/2\right) \left( rb^{2}-rb^{\dagger 2}\right)}$. The instantaneous state of system is
 \begin{align}
 |\Psi(t) \rangle =&S^{\dagger }\left( r\right) e^{-i\omega
  _{s}tb^{\dagger }b}S\left( r\right) |\beta \rangle, \nonumber\\
 =&S^{\dagger }\left( r\right) S\left( r'\right) |\beta(t) \rangle,
 \end{align}%
 where $S\left(r'\right) =e^{\left(1/2\right) \left( r'^{\ast }b^{2}-r'b^{\dagger 2}\right) }$ with $r'
 =re^{-2i\omega _{s}t}
 $, and $\beta(t)=e^{-i\omega_st}\beta$. To visualize the periodic squeezing behavior, we illustrated the instantaneous squeezing degree of mechanical mode in Fig.\,2(a) of the main text. The detailed derivation is shown in the following. Instantaneous mean values of $\langle X\rangle^2$ and $\langle X^2\rangle$ with $X=(b+b^{\dag})/\sqrt{2}$ being a dimensionless position operator, are
 \begin{align}
 \langle\Psi(t)|X|\Psi(t)\rangle^2={2}e^{2r}\left(\cosh ^2r{\rm Re}[\beta(t)]^2-2\sinh r\cosh r{\rm Re}[\beta(t)] {\rm Re}[(\beta(t)e^{2i\omega_st})]+\sinh ^2 r{\rm Re}[\beta(t)e^{2i\omega_st}]^2\right),
 \end{align}
 \begin{eqnarray}
 \langle\Psi(t)|X^2|\Psi(t)\rangle^2={e^{2r}}\left(\begin{array}{c}2({\rm Re}[\beta(t)])^2\cosh^2 r+2{\rm Re}[\beta(t) e^{2i\omega_st}]^2\sinh^2 r-4{\rm Re}[\beta(t) e^{2i\omega_st}]{\rm Re}[\beta(t)]\sinh r\cosh r\\+\frac{1}{2}(\sinh ^2 r+\cosh^2 r)-\frac{1}{2}(e^{-2i\omega_st}+e^{2i\omega_st})\sinh r\cosh r\end{array}\right).\nonumber\\
 \end{eqnarray}
 Then we obtain the variance of $X$
 \begin{align}
 \langle \delta ^2X(t)\rangle&=\langle X(t)^2\rangle-\langle X(t)\rangle^2
 =\frac{e^{2r}}{2}\left[\cosh (2r)-\cos(2\omega_st)\sinh (2r)\right].
 \end{align}
 Fig.\,\ref{figS3} clearly shows that $\langle \delta ^2X(t)\rangle$ evolves with a period $T=\pi/\omega_s$. It reaches the minimum $\langle \delta X^2\rangle_{\rm min}=1/2$ at  $t=m\pi/\omega_s(m=0,1,2,...)$ and the maximum $\langle \delta X^2\rangle_{\rm max}=e^{4r}/2$ at $t=m\pi/(2\omega_s)(m=1,2,3,...)$. For a certain squeezing parameter $r$, we define the instantaneous squeezing degree $S(t)=10\log_{10}\left({\langle \delta ^2X(t)\rangle}/{\langle \delta ^2 X\rangle_{\rm min}}\right){\rm dB}$, and its maximum value is
	\begin{align}S_{\rm max}&=10\log_{10}\left(\frac{\langle \delta ^2X\rangle_{\rm max}}{\langle \delta ^2 X\rangle_{\rm min}}\right){\rm dB},\nonumber\\
&=10\log_{10}(e^{4r})\,{\rm dB}.\end{align}

\section{State tomography}\label{section4}
As discussed in the main text, the well-defined displaced phase-accumulation-efficiency $\widetilde{\mathcal{P}}_n$ is experimentally observable. Specifically, by projecting the system density matrix $\rho_c(\tau_1)$ on the basis vectors $\{|\!\uparrow\rangle, |\!\downarrow\rangle\}$, where
$|\!\uparrow\rangle=(1/\sqrt{2})(|0\rangle+|l\rangle)$, $|\!\downarrow\rangle=(1/\sqrt{2})(|0\rangle-|l\rangle)$ with Fock state of cavity mode $|l\rangle$, we obtain
	\begin{align}
	\rho_{\downarrow\downarrow}(\tau_1)&=\frac{1}{2}(\langle0|-\langle l|)\left(e^{-|\alpha|^2}\sum_{n,n'}\frac{\alpha^n\alpha^{*n'}}{\sqrt{n!n'!}}e^{i(\Phi_n(\tau_1)-\Phi_{n'}(\tau_1))}|n\rangle\langle n'|\right)(|0\rangle-|l\rangle),\nonumber\\
	&=\frac{e^{-|\alpha|^2}}{2}\left[1+\frac{\alpha^{2l}}{l!}-\frac{\alpha^l}{\sqrt{l!}}\left(e^{i\Phi_0(\tau_1)-i\Phi_l(\tau_1)}+e^{i\Phi_l(\tau_1)-i\Phi_0(\tau_1)}\right)\right],\nonumber\\
	&=\frac{e^{-|\alpha|^2}}{2}\left[1+\frac{\alpha^{2l}}{l!}-\frac{\alpha^l}{\sqrt{l!}}2\cos\left(2\pi(\tilde{\lambda}_s^2l^2-2\tilde{f}_s\tilde{\lambda}_sl)\right)\right].\end{align}
	As above we can obtain the other elements
	\begin{align}
	\rho_{\uparrow\uparrow}(\tau_1)&=\frac{1}{2}(\langle0|+\langle l|)\left(e^{-|\alpha|^2}\sum_{n,n'}\frac{\alpha^n\alpha^{*n'}}{\sqrt{n!n'!}}e^{i(\Phi_n(\tau_1)-\Phi_{n'}(\tau_1))}|n\rangle\langle n'|\right)(|0\rangle+|l\rangle),\nonumber\\
	&=\frac{e^{-|\alpha|^2}}{2}\left[1+\frac{\alpha^{2l}}{l!}+\frac{\alpha^l}{\sqrt{l!}}2\cos\left(2\pi(\tilde{\lambda}_s^2l^2-2\tilde{f}_s\tilde{\lambda}_sl)\right)\right].\end{align}
	\begin{align}
	\rho_{\downarrow\uparrow}(\tau_1)&=\frac{1}{2}(\langle0|-\langle l|)\left(e^{-|\alpha|^2}\sum_{n,n'}\frac{\alpha^n\alpha^{*n'}}{\sqrt{n!n'!}}e^{i(\Phi_n(\tau_1)-\Phi_{n'}(\tau_1))}|n\rangle\langle n'|\right)(|0\rangle+|l\rangle),\nonumber\\
	&=\frac{e^{-|\alpha|^2}}{2}\left[1-\frac{\alpha^{2l}}{l!}+\frac{\alpha^l}{\sqrt{l!}}2i\sin\left(2\pi(\tilde{\lambda}_s^2l^2-2\tilde{f}_s\tilde{\lambda}_sl)\right)\right].\end{align}
	\begin{align}
	\rho_{\uparrow\downarrow}(\tau_1)&=\frac{1}{2}(\langle0|-\langle l|)\left(e^{-|\alpha|^2}\sum_{n,n'}\frac{\alpha^n\alpha^{*n'}}{\sqrt{n!n'!}}e^{i(\Phi_n(\tau_1)-\Phi_{n'}(\tau_1))}|n\rangle\langle n'|\right)(|0\rangle+|l\rangle),\nonumber\\
	&=\frac{e^{-|\alpha|^2}}{2}\left[1-\frac{\alpha^{2l}}{l!}-\frac{\alpha^l}{\sqrt{l!}}2i\sin\left(2\pi(\tilde{\lambda}_s^2l^2-2\tilde{f}_s\tilde{\lambda}_sl)\right)\right].
	\end{align}These matrix elements are shown in Fig.\,2(b) of the main text. The difference of diagonal elements is
\begin{align}
	\sigma_z(\tau_1)=\rho_{\uparrow\uparrow}(\tau_1)-\rho_{\downarrow\downarrow}(\tau_1)=2\,e^{-|\alpha|^2}\frac{\alpha^l}{\sqrt{l!}}\cos\left[\Phi_l(\tau_1)-\Phi_0(\tau_1)\right].
\end{align}
Evidently, we can obtain the phase difference $\Delta\Phi_l(\tau_1)=\Phi_l(\tau_1)-\Phi_0(\tau_1)$ by directly measuring $\rho_{\downarrow\downarrow}(\tau_1)$ and $\rho_{\uparrow\uparrow}(\tau_1)$, and then  $\widetilde{\mathcal{P}}_l=\Delta \Phi_l(\tau_1)/\tau_1$ is experimentally observable.

\section{Quantum Fisher information and ultimate bound of sensitivity}\label{section5}
\begin{figure}[h]
\includegraphics[width=13.4cm]{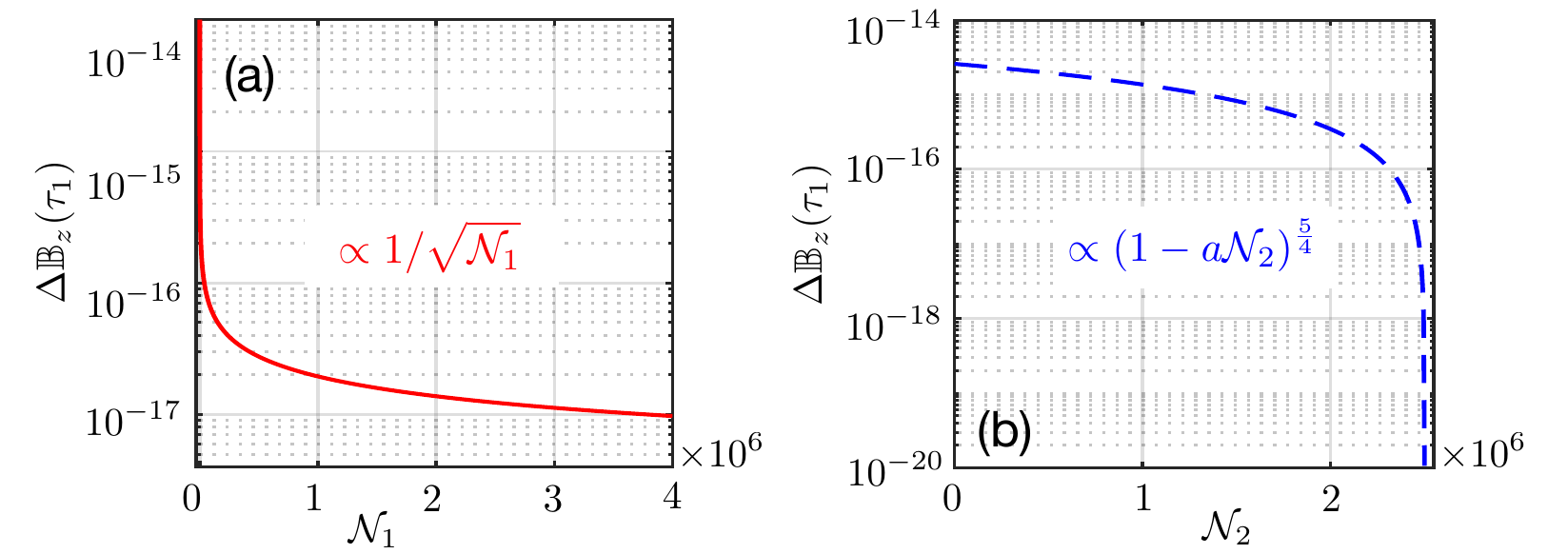}
\caption{Ultimate bound of sensitivity $\Delta \mathbb{B}_z(\tau_1)$ in units of T/${\rm \sqrt{Hz}}$ versus the resources $\mathcal{N}_1$ and $\mathcal{N}_2$. We considered  $\mathcal{N}_2=2.45\times 10^6$ in (a), $\mathcal{N}_1=10^6$ and $a$ is a constant in (b). Other parameters are same as the main text.}
	\label{figS4}
\end{figure}

In this section, we will present the detailed derivation of QFI in our model. From Eq.\,(\ref{EQ:Psitau}), we can obtain the state at time $\tau_1$ reads
\begin{align}\label{EQ:Psitau1}
|\Psi(\tau_1)\rangle&=e^{-|\alpha|^2/2}\sum_{n}\frac{\alpha^n}{\sqrt{n!}}e^{i2\pi(\tilde{\lambda}_sn-\tilde{f}_s)^2}|n\rangle|\beta\rangle.
\end{align}
As discussed before, the optical and mechanical modes absolutely decoupled at time $\tau_1$, and all information about $B_z$ is transduced into optical phase $\Phi_n(\tau_1)$. Starting from the system state $|\Psi(\tau_1)\rangle$, the QFI is given by\,\cite{PARIS2009,Toth_2014,Liu_2019}
	\begin{align}\label{EQ:Qtau}
	\mathcal{F}_{q}({\tau_1})=4\left(\langle \partial _{B_z}\Psi(\tau_1)|\partial_{B_z}\Psi(\tau_1)\rangle-|\langle \Psi(\tau_1)|\partial_{B_z}\Psi(\tau_1)\rangle|^2\right).
	\end{align}
	Taking the partial derivative of Eq.\,(\ref{EQ:Psitau1}) with respect to $B_z$, we obtain
	\begin{align}
	\partial_{B_z}|\Psi(\tau_1)\rangle=\frac{\partial |\Psi(\tau_1)\rangle}{\partial B_z}=-i4\pi c_{\rm act}\sqrt{\frac{1}{2m\hbar\omega_m^3}}e^{3r}e^{-|\alpha|^2}\sum_{n}\frac{\alpha^n}{\sqrt{n!}}(\tilde{\lambda}_sn-\tilde{f}_s)e^{i2\pi(\tilde{\lambda}_sn-\tilde{f}_s)^2}|n\rangle|\beta\rangle,
	\end{align}
	and
	\begin{align}\label{EQ:PsiBz}
	\langle\partial_{B_z}\Psi(\tau_1)|\partial_{B_z}\Psi(\tau_1)\rangle&=\frac{8\pi^2c_{\rm act}^2e^{6r}}{m\hbar\omega_m^3}e^{-|\alpha|^2}\left(\sum_n\frac{(\tilde{\lambda}_sn-\tilde{f}_s)^2\alpha^{2n}}{n!}\right),\nonumber\\
	&=\frac{8\pi^2c_{\rm act}^2e^{6r}}{m\hbar\omega_m^3}\left[\tilde{\lambda}_s^2\alpha^4+(\tilde{\lambda}_s^2-2\tilde{\lambda}_s\tilde{f}_s)\alpha^2+\tilde{f}_s^2\right].
	\end{align}
	In derivating Eq.\,(\ref{EQ:PsiBz}), we have used the following Taylor series expansion
	\begin{align}
	&\sum_{n}{\frac{n^2\alpha^{2n}}{n!}}=\alpha^2(1+\alpha^2)e^{\alpha^2},\\
	&\sum_{n}\frac{n\alpha^{2n}}{n!}=\alpha^2e^{\alpha^2},\\
	&\sum_n\frac{\alpha^{2n}}{n!}=e^{\alpha^2}.
	\end{align}
	Following the same method introduced above, we can obtain
	\begin{align}\label{EQ:PsitauBz}
	|\langle \Psi(\tau_1)|\partial_{B_z}\Psi(\tau_1)|^2=\frac{8\pi^2c_{\rm act}^2e^{6r}}{m\hbar\omega_m^3}(\tilde{\lambda}_s^2\alpha^4-2\tilde{\lambda}_s\tilde{f}_s\alpha^2+\tilde{f}_s^2).
	\end{align}
	Substituting Eqs.\,(\ref{EQ:PsiBz}) and (\ref{EQ:PsitauBz}) to Eq.\,(\ref{EQ:Qtau}) and using $c_{\rm act}=m\omega_m^2L{\alpha_{\rm mag}/}{E}$, we obtain
	\begin{align}\label{EQ:Qt}
	\mathcal{F}_q(\tau_1)=\frac{32\pi^2m\lambda_1^2L^2\alpha_{\rm mag}^2}{\hbar\omega_mE^2}\mathcal{N}_1e^{12r},
	\end{align}
	where $\mathcal{N}_1=\alpha^2$.
	To clearly see the dependence of QFI on the mean photon number, we do the following expansion
	\begin{align}
	e^{12r}&=\frac{1}{\left[1-{(4\lambda_2/\omega_m)\mathcal{N}_2}\right]^3},
	\end{align}
	where $\mathcal{N}_2=|\xi|^2$. Then $\mathcal{F}_q(\tau_1)$ can also be expressed as
	\begin{align}\label{EQ:Qtau2}
	\mathcal{F}_q(\tau_1)=\frac{32\pi^2m\lambda_1^2L^2\alpha_{\rm mag}^2}{\hbar\omega_mE^2}\frac{\mathcal{N}_1}{\left[1-{(4\lambda_2/\omega_m)\mathcal{N}_2}\right]^3}.
	\end{align}
According to the Cram\'{e}r-Rao inequality, the QFI gives the ultimate lower limit of parameter estimation. Obtaining this lower bound requires the adoption of optimal measurements for the system, but it does not reveal which specific measurement is required to achieve it. The ultimate bound of sensitivity is given by
\begin{align}
\Delta\mathbb{B}_z(\tau_1)&=\frac{Ee^{-5r}}{4\pi\lambda_1L\alpha_{\rm mag}}\sqrt{\frac{\pi\hbar}{m\mathcal{N}_1}}({\rm T/\sqrt{Hz}}),\nonumber\\
&=\frac{E}{4\pi\lambda_1L\alpha_{\rm mag}}\sqrt{\frac{\pi\hbar[1-(4\lambda_2/\omega_m)\mathcal{N}_2]^\frac{5}{2}}{m\mathcal{N}_1}}({\rm T/\sqrt{Hz}}).
\end{align}
Fig.\,\ref{figS4} clearly shows the different scaling of $\Delta \mathbb{B}_z(\tau_1)$ with respect to the resources $\mathcal{N}_1$
and $\mathcal{N}_2$. Evidently, resources $\mathcal{N}_1$ and $\mathcal{N}_2$ jointly determine the ultimate sensitivity.

\section{CFI and specific sensitivity}\label{section6}
\subsection{CFI without dissipation}
In this section, let us offer the detailed derivation of the CFI, i.e., Eq.\,(6) used in the main text. Normally, the CFI, corresponding to a specific measurement of $B_z$,  is given by
\begin{align}\label{EQFI}
\mathcal{F}_c(t)=\int \frac{1}{P(X|B_z)}\left(\frac{\partial P(X|B_z)}{\partial B_z}\right)^2{\rm d}X.
\end{align}
where $P(X|B_z)$ represents the conditional probability of measuring $X$ relied on parameter $B_z$. With the positive-operator valued measure elements $\{\Pi_X\}$, we obtain $P(X|B_z)={\rm Tr}[\Pi_X \rho(t)]$. Here we consider a general homodyne measurement on the traced-out cavity state $\rho_c(t)$ with the observable operator $X_{\theta}=(a_1e^{-i\theta}+a_1^{\dag}e^{i\theta})/\sqrt{2}$, where $\theta$ is the phase of local oscillator. The cases of $\theta=0$ and $\theta=\pi/2$ correspond to the position and momentum measurements, respectively. Then the conditional probability becomes
	\begin{align}\label{EQPXB}
	P(X_{\theta}|B_z)={\rm Tr}[|X_{\theta}\rangle\langle X_{\theta}|\rho_c(t)],
	\end{align}
where $|X_{\theta}\rangle$ is the eigenstate of $X_{\theta}$. Using the inner product $\langle n|X_{\theta}\rangle=\pi^{-1/4}[2^{n}(n!)]^{-1/2}\exp(-X_{\theta}^2/2)H_n(X_{\theta})\exp(in\theta)$ with $H_n(X_\theta)=\exp(X_\theta^2/2)\left(X_\theta-{\rm d}/{\rm d}X_\theta\right)^n\exp(-X_\theta^2/2)$ being the Hermite polynomials of order $n$, we can rewrite Eq.\,(\ref{EQPXB}) as
	\begin{align}\label{EQPXBz}
	P(X_{\theta}|B_z)&=e^{-|\alpha|^2}\sum_{n,n'}\left[\frac{\alpha^n(\alpha^*)^{n'}}{\sqrt{n!n'!}}e^{i\left[\tilde{\lambda}_s^2(n^2-n'^2)-2\tilde{\lambda}_s\tilde{f}_s(n-n')\right](\omega_st-\sin\omega_st)}\frac{e^{-X_{\theta}^2}}{\sqrt{\pi}}\frac{H_n(X_{\theta})H_{n'}(X_{\theta})e^{-i\theta(n-n')}}{2^{(n+n')/2}\sqrt{n!n'!}}\right.\nonumber\\
	&\left.\times e^{i\tilde{\lambda}_s(n-n')[e^{-r}\beta_{\rm Re}\sin \omega_st-e^{r}\beta_{\rm Im}(\cos \omega_st-1)]}e^{-(|\varphi_n|^2+|\varphi_{n'}|^2)/2+\varphi_{n'}^*\varphi_n}\right].
	\end{align}
	Substituting Eq.\,(\ref{EQPXBz}) into Eq.\,(\ref{EQFI}), we obtain the analytical expression of CFI at time  $\tau_1=2\pi/\omega_s$ is
	\begin{align}
	\mathcal{F}_c(\tau_1)=\frac{8\pi^2m\lambda_1^2L^2\alpha_{\rm mag}^2}{\hbar \omega_mE^2}e^{12r}[2\sin\theta\alpha_{\rm Re}-2\cos\theta\alpha_{\rm Im}]^2,
	\end{align}
	and it can be reduced further as
	\begin{equation}
	\mathcal{F}_c(\tau_1)=\begin{dcases}
	\frac{32\pi^2m\lambda_1^2L^2\alpha_{\rm mag}^2|\alpha_{\rm Im}|^2}{\hbar \omega_mE^2}e^{12r}, &{\rm for}\,\,\,\theta=0,\\
	\frac{32\pi^2m\lambda_1^2L^2\alpha_{\rm mag}^2|\alpha_{\rm Re}|^2}{\hbar \omega_mE^2}e^{12r}, &{\rm for} \,\,\,\theta=\pi/2.
	\end{dcases}
	\end{equation}
\begin{figure}[h]
	\includegraphics[width=13.2cm]{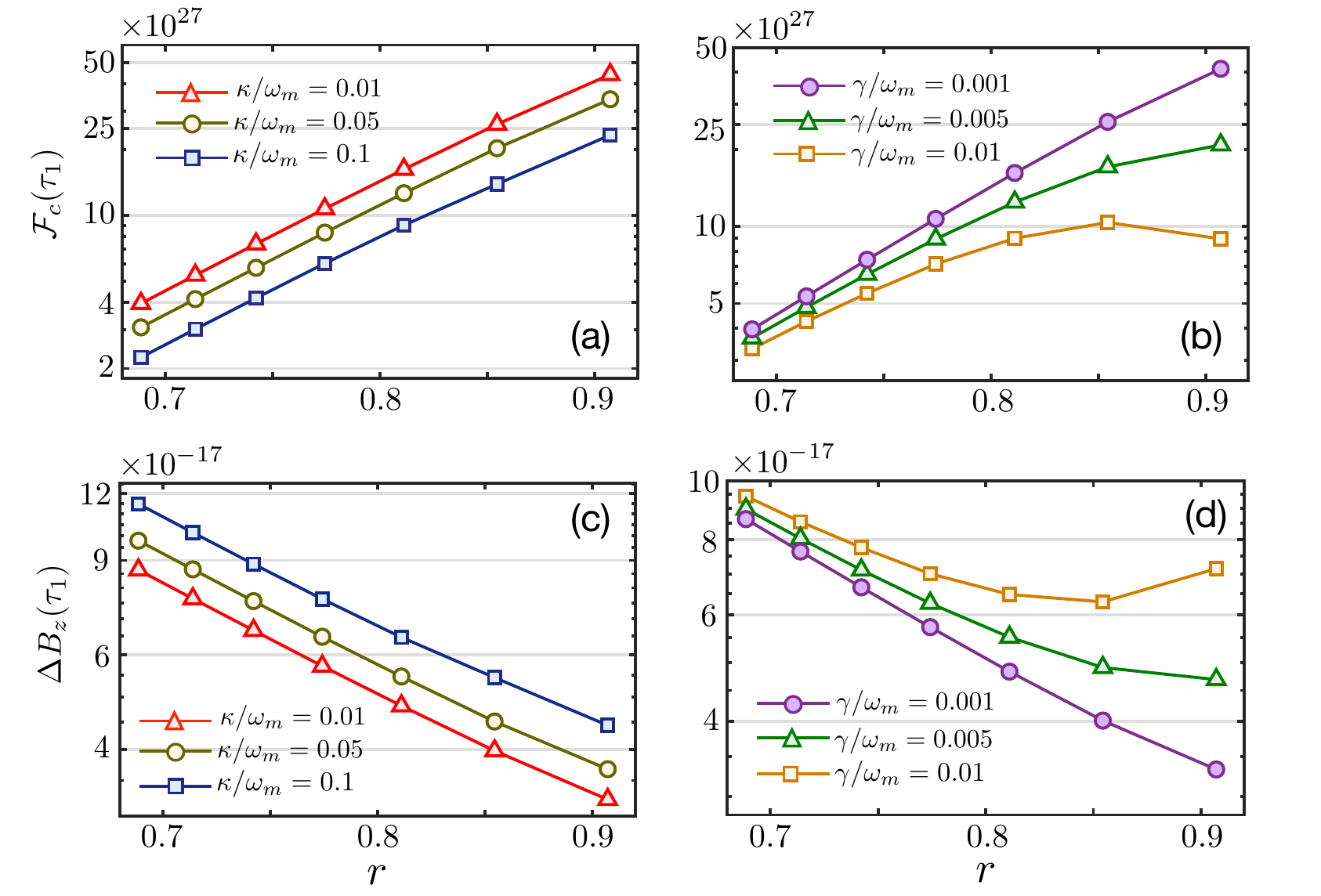}
	\caption{ Classical Fisher information $\mathcal{F}_c(\tau_1)$ versus squeezing parameter $r$ for different (a) cavity decay rates $\kappa$, and (b) mechanical decay rates $\gamma$. In (c,d), we show the corresponding sensitivity limit $\Delta B_z (\tau_1)$ in units of T/${\rm \sqrt{Hz}}$. We have chosen the parameters as (a,c) $\gamma/\omega_m=0.001$, and (b,d) $\kappa/\omega_m=0.01$. Other parameters are $ \omega_m=2\pi\times 134\,{\rm kHz}, f=0.01\omega_m$ and $\bar{n}_{\rm th}=10$.}
	\label{figS5}
\end{figure}

\subsection{CFI and specific sensitivity limit with system dissipation}
In this subsection, let us present the detailed derivation for the CFI in the case of including system dissipation. In a practical experimental setup, the dissipation caused by the system-bath coupling should be considered. Then the full dynamics of the system satisfy the following master equation
	\begin{align}\label{EQ:rho1}
	\frac{\partial\rho(t)}{\partial t}=&-\frac{i}{\hbar}[H,\rho(t)]+\frac{\kappa}{2}\left[2a_1\rho a_1^{\dag}-\rho a_1^{\dag}a_1-a_1^{\dag}a_1\rho\right]+\frac{\gamma}{2}(\bar{n}_{\rm th}+1)\left[2b\rho b^{\dag}-\rho b^{\dag}b-b^{\dag}b\rho\right]\nonumber\\
	&+\frac{\gamma}{2}\bar{n}_{\rm th}\left[2b^{\dag}\rho b-\rho bb^{\dag}-bb^{\dag}\rho\right],
	\end{align}
	where $\kappa(\gamma)$ is the cavity (mechanical) decay rate, and $\bar{n}_{\rm th}$ is the thermal phonon number of mechanical mode. Hamiltonian $H$ is shown in Eq.\,(\ref{EQ:H}). Applying a squeezing transformation to Eq.\,(\ref{EQ:rho1}) with $S(r)=\exp[r(b^2-b^{\dag 2})/2]$ and $r=-(1/4)\ln(1-4\lambda_2\mathcal{N}_2/\omega_m)$, we obtain	
	\begin{eqnarray}\label{rho}
	\frac{\partial{\rho}_s(t)}{\partial t}
	&=&-\frac{i}{\hbar }\left[ H_s,\rho_s\right] +\frac{{\kappa}}{2}\left[ 2a_1\rho_sa_1^{\dagger }-\rho
	_sa_1^{\dagger }a_1-a_1^{\dagger }a_1\rho_s\right]\nonumber\\&&+{\gamma}(\bar{n}_{\rm th}+1)\left(
	{\mathcal D}[b]\rho_s\cosh^2 r+{\mathcal D}[b^{\dag}]\rho_s\sinh^2 r+{\mathcal G}[b]\rho_s \sinh r\cosh r+{\mathcal G}[b^{\dag}]\rho_s \sinh r\cosh r
	\right)\nonumber\\
	&&+{\gamma}\bar{n}_{\rm th}\left(
	\mathcal
	D[b^{\dag}]\rho_s\cosh^2 r+{\mathcal D}[b]\rho_s\sinh^2 r+{\mathcal G}[b]\rho_s \sinh r\cosh r+{\mathcal G}[b^{\dag}]\rho_s \sinh r\cosh r
	\right),
	\end{eqnarray}
where $\rho_s=S(r)\rho S^{\dag}(r)$, $H_s=S(r)HS^{\dag}(r)$ and the Lindblad superoperators read
	\begin{align}
	\mathcal{D}[o]\rho=o\rho o^{\dag}-(o^{\dag}o\rho+\rho o^{\dag}o)/2,\,\,\,\,\,\,\,\,\,
	\mathcal{G}[o]\rho=o\rho o-(oo\rho+\rho oo)/2.
	\end{align}
We can numerically calculate $\rho$ by $\rho=S^{\dag}\rho_s S(r)$. Once we numerically obtain the density matrix, we can further calculate the CFI.

Next we discuss the effect of system dissipations on the CFI and specific  sensitivity limit $\Delta B_z(\tau_1)$. By numerically solving the master equation \,(\ref{rho})\,\cite{Johansson201201}, we can obtain the CFI at time $\mathcal{F}_c(\tau_1)$ and the sensitivity $\Delta B_z(\tau_1)$ in the presence of system dissipations. In addition to the main results shown in the main text, here we supplement some numerical results indicating the influences of squeezing parameter $r$ and system decay rates on the CFI and sensitivity in Figs.\,\ref{figS5}.  It is shown that, for a certain squeezing parameter $r$, the increased cavity decay $\kappa$ leads to a decline of CFI $\mathcal{F}_{c}(\tau_1)$. In spite of this, its corresponding sensitivities still maintain at the order of $10^{-17}\,{\rm T/\sqrt{Hz}}$ [see Fig.\,\ref{figS5}(c)].
%It is shown in Figs.\,\ref{figS5} that the mechanical squeezing could enhance the CFI but decrease the time window of implementing magnetic detection with high precision [This is consistent with the conclusion we obtained in the Sec.\,\ref{section6}B].

In Fig.\,\ref{figS5}(b) we plot $\mathcal{F}_c(\tau_1)$ varying with squeezing parameter $r$, for different mechanical decays. Specifically, for a small squeezing parameter, mechanical dissipation exerts little influence on the CFI. The increasing squeezing parameter amplifies the noise coming from the mechanical bath which causes the influence of mechanical decay $\gamma$ becomes larger. This is definitely different from the effect of optical decay $\kappa$ exerted on the CFI [see Fig.\,\ref{figS5}(a)]. Even though the mechanical decay reduces achievable CFI, numerically simulation has clearly shown that the measurement maintains a high accuracy in the presence of dissipation [see Figs.\,\ref{figS5}(d)]. In other words, the measurement sensitivity of our proposal is insensitive to the system dissipation due to the mechanical oscillator and optical cavity decoupled at detection time $\tau_1$.

{\Tin{\section{Frequency response of resonant dual-coupling magnetometer}\label{section7}
In the main text, we proposed a scheme to measure the dc magnetic fields (or static magnetic fields) in a dual-coupling optomechanical system. We calculated the quantum and classical Fisher information $F_q$ and $F_c$, and then obtained the fundamental bound to the sensitivity $\Delta B_z$. This sensitivity gives the theoretical lower limit of the measurement precision based on our proposal, which is several orders of magnitude lower than that achieved in recent experiments\,\cite{fforstner201201, Li2021, ccolombano2020}.
Note that, besides detecting a static magnetic field discussed in the main text, in principle, our proposal can also work as a resonant sensor to be applied to detect the alternating magnetic fields. In the following, we will estimate the frequency response characteristics including the bandwidth issue, when the proposed dual-optomechanical system is used to detect the alternating magnetic fields.

\begin{figure}[h]
\includegraphics[width=15.0cm]{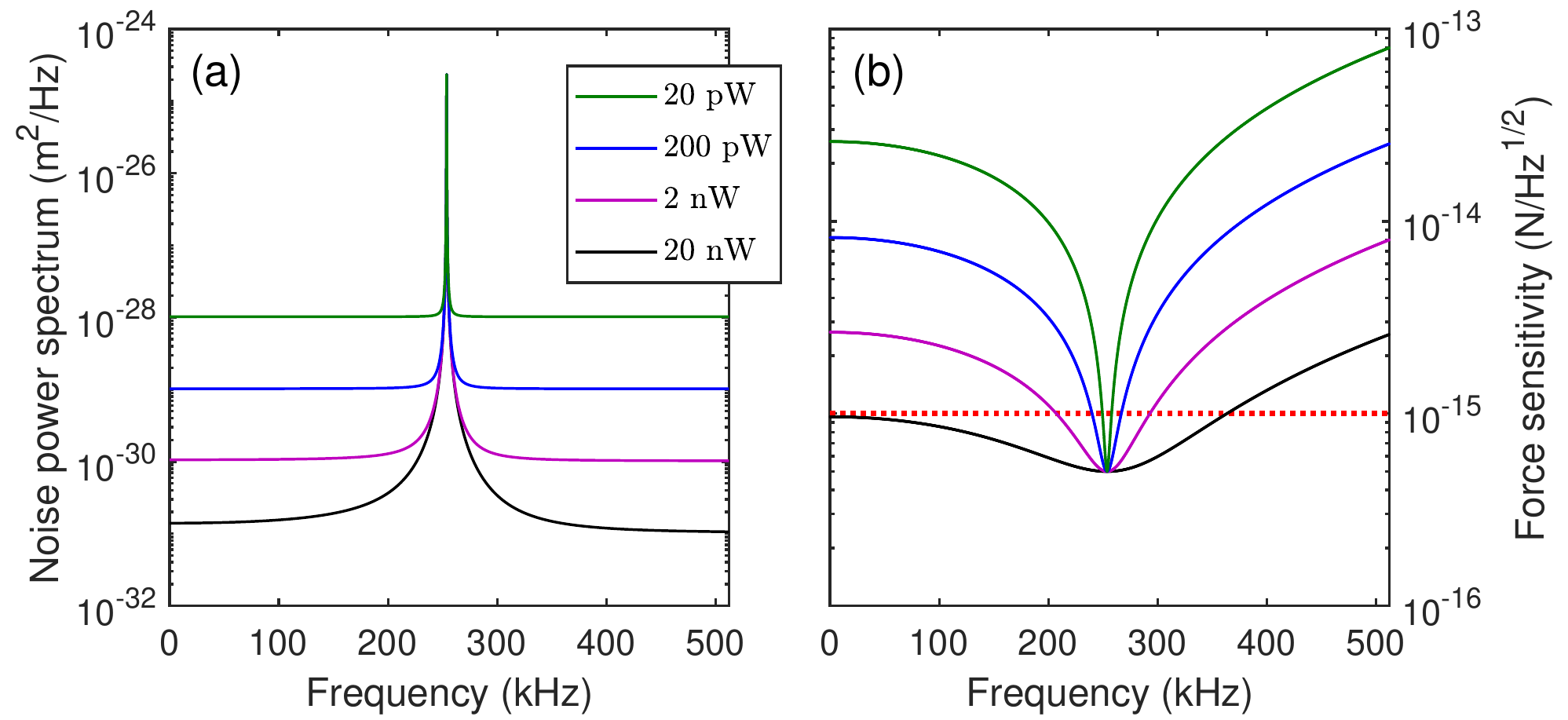}
\caption{{\Tin{(a) Displacement thermal noise power spectra $S_{xx}(\omega)$ as a function of frequency $\omega$, for different values of powers, i.e., $P=20\, {\rm pW}$ (green), $200\, {\rm pW} $ (blue), $2\,{\rm nW}$ (magenta), and $20\,{\rm nW}$ (black). (b) Force sensitivity $\sqrt{S_{FF}(\omega)}$ vs frequency. Plotting these figures, we referred to the parameters used in optomechanical experiments\,\cite{Thompson2008,Li2021}, i.e., $\omega_m =2\pi\times1.34\times10^5$Hz, $\gamma = 2\pi\times 0.12$ Hz, $m=4\times 10^{-11}$ kg, $\omega_c=\omega_L=2\pi\times 10^{14}\,$Hz, $\kappa= 2\pi\times10^8$ Hz, $\kappa_{\rm ex}= \kappa$, the optomechanical coupling $G = 500$ \,MHz/nm,  $\eta=0.8$, $T=300 \,{\rm K}$, and $r = 0.6$.}}}
	\label{figS6}
\end{figure}

In cavity optomechanical magnetometers, the magnetostrictive material expands, exerting a force upon the mechanical oscillator. The mechanical motion modulates the optical cavity field via the radiation-pressure coupling. Meanwhile, the phase shift of the mechanical motions encoded with magnetic signal is transferred to the optical field, which enables us to optically readout mechanical motions. The combination of mechanical and optical resonances provides enhanced mechanical response to applied forces and optical readout with high precision. Experimentally, the sensitivity of such a magnetometer is defined as the minimum detectable signal, and can be quantified by the noise equivalent signal. Thus, the performance of the optomechanical magnetometer relies on how efficiently the magnetic signal drives the mechanical modes above the experiment noise level.
Generally speaking, the noise sources in the optomechanical systems mainly consist of the thermal noise from the thermal environment with nonzero temperature and the photon shot noise from the probe laser.

For simplicity, we consider a single mode of mechanical resonance, whose response to an external force in the
frequency domain is quantified by the mechanical susceptibility of the resonator $\chi(\omega)=1/(m(\omega_s^2-\omega^2-i\omega\gamma))$ with effective mass  $m$, frequency $\omega_s$ and mechanical decay rate $\gamma$.
The magnetometer sensor is essentially a force sensor, whose sensitivity of displacement at the frequency $\omega$ is determined by $\sqrt{S_{xx}(\omega)}$, where the noise spectrum is defined as $S_{xx}(\omega)=\int_{-\infty}^{+\infty}\langle x(t)x(0)\rangle e^{i\omega t}dt$.
According to the fluctuation dissipation theorem,  a mechanical resonator with frequency $\omega_s$ experiences a thermal noise force $F_{\rm th}=\sqrt{2m\gamma k_BT}$\,\cite{Aspelmeyer2014} at a temperature $T$, and the corresponding thermal force power spectrum reads\,\cite{Li2021}
\begin{align}
S_{xx}^{\rm th}=\frac{2\gamma k_BT}{m[(\omega_s^2-\omega^2)^2-\omega^2\gamma^2]}.
\end{align}
The displacement noise power spectrum from the laser shot noise,
\begin{align}
S_{xx}^{\rm shot}(\omega)=\frac{\kappa}{16\eta NG^2}(1+\frac{\omega^2}{\kappa^2}),\label{EQSXXshot}
\end{align}
where $N=4P\kappa_{\rm ex}/(\hbar\omega_L\kappa^2)$ is the intracavity photon number\,\cite{Aspelmeyer2014}, $\omega_L$ and $P$ are the frequency and power of the probe field, respectively. Here $\kappa$ is the total cavity loss rate, $\kappa_{\rm ex}$ refers to the loss rate associated with the input coupling, $\eta$ is the optical detection efficiency, and $G={\rm d}\omega_c/{\rm d}t$ is the optomechanical coupling strength.

In Fig.\,\ref{figS6}, we numerically simulate the displacement noise power spectrum and the corresponding force sensitivity at different probe powers. It can be seen from Fig.\,\ref{figS6}(a) that these noise power spectra show a peak at resonance, i.e., $\omega=\omega_s$. With the increase of probe power, the shot noise is suppressed effectively [see Eq.\,(\ref{EQSXXshot})]. Meanwhile, the peak of the force sensitivity also occurs at resonance [ses Fig.\,\ref{figS6}(b)], which is determined by the mechanical decay rate $\gamma$. The red dotted curve in Fig.\,\ref{figS6}(b) denotes the frequency range in which the sensitivity is better than twice of the peak sensitivity, which is defined as the bandwidth\,\cite{Li2021}. It can be seen that, the larger the probe power is, the broader bandwidth is. For example, when the probe power is $20 \,{\rm nW}$, the thermal-noise-limited frequency range covers $0\sim 380$ kHz with the central resonant frequency 250 kHz. Therefore, similar as a general optomechanical system\,\cite{Li2021}, here the dual-coupling magnetometer also has a broad bandwidth, when it is used as a resonant sensor.}}

%\vspace{8mm} \small{$^*$Corresponding author.

%$~$$\,$jinghui73@foxmail.com}

%\bibliography{Sreferences}


\end{document}